\documentclass[aps,prb,twocolumn,amsmath,amssymb,floatfix,groupedaddress,superscriptaddress,longbibliography]{revtex4-2}
\usepackage{graphicx}
\usepackage[english]{babel}
\usepackage{times}
\usepackage{dcolumn}
\usepackage[usenames,dvipsnames]{color}
\usepackage{units}
\usepackage{color}
\usepackage[utf8]{inputenc}

\begin{document}

\title{Yu-Shiba-Rusinov states, the BCS-BEC crossover, and the exact solution in the flat-band limit}

\author{R. \v{Z}itko}

\affiliation{Jo\v{z}ef Stefan Institute, Jamova 39, SI-1000 Ljubljana,
Slovenia}
\affiliation{Faculty  of Mathematics and Physics, University of
Ljubljana, Jadranska 19, SI-1000 Ljubljana, Slovenia}

\author{L. Pave\v{s}i\'c}

\affiliation{Jo\v{z}ef Stefan Institute, Jamova 39, SI-1000 Ljubljana,
Slovenia}
\affiliation{Faculty  of Mathematics and Physics, University of
Ljubljana, Jadranska 19, SI-1000 Ljubljana, Slovenia}

\begin{abstract}
We study the subgap Yu-Shiba-Rusinov (YSR) states in
the Richardson's model of a superconductor with a magnetic impurity
for different electron pairing strengths from the weak-coupling
Bardeen-Cooper-Schrieffer (BCS) regime to the strong-coupling
Bose-Einstein-condensate (BEC) regime. We observe that the effect of
the increasing pairing strength on the YSR excitation spectrum as a
function of hybridisation strength and impurity on-site
potential is only quantitative and, in fact, rather weak when the
results are appropriately rescaled. We furthermore show that the
problem is analytically solvable in the deep BEC limit which is equivalent to
flat-band superconductivity. This exact solution can be related to a
zero-bandwidth (ZBW) effective BCS mean-field Hamiltonian where the superconductor is
described by a single electron level with onsite pairing.  The small difference between
the BCS and BEC regimes of the Richardson's model  explains
the success of the simple ZBW calculations for BCS mean-field
Hamiltonians. A ZBW model requires only a suitable parameter rescaling to become useful as a quantitative predictive tool for the full problem.
\end{abstract}

\maketitle

\newcommand{\vc}[1]{{\mathbf{#1}}}
\newcommand{\vck}{\vc{k}}
\newcommand{\braket}[2]{\langle#1|#2\rangle}
\newcommand{\expv}[1]{\langle #1 \rangle}
\newcommand{\corr}[1]{\langle\langle #1 \rangle\rangle}
\newcommand{\bra}[1]{\langle #1 |}
\newcommand{\ket}[1]{| #1 \rangle}
\newcommand{\Tr}{\mathrm{Tr}}
\newcommand{\kor}[1]{\langle\langle #1 \rangle\rangle}
\newcommand{\degg}{^\circ}
\renewcommand{\Im}{\mathrm{Im}\,}
\renewcommand{\Re}{\mathrm{Re}\,}
\newcommand{\dtN}{{\dot N}}
\newcommand{\dtQ}{{\dot Q}}
\newcommand{\sgn}{\mathrm{sgn}}
\newcommand{\beq}[1]{\begin{equation} #1 \end{equation}}
\newcommand{\beqz}[1]{\begin{equation*} #1 \end{equation*}}
\newcommand{\one}{\mathbf{1}}

\newcommand{\dd}{\mathrm{d}}
\newcommand{\Res}{\mathrm{Res}}
\newcommand{\sign}{\mathrm{sign}}

\newcommand{\imp}{\mathrm{imp}}
\newcommand{\Simp}{\mathbf{S}_\mathrm{imp}}
\newcommand{\bfsc}{\mathbf{s}_c}

\newcommand{\HQD}{H_\mathrm{QD}}
\newcommand{\nimp}{n_\mathrm{imp}}
\newcommand{\epsQD}{\epsilon}
\newcommand{\UQD}{U}

\newcommand{\up}{\uparrow}
\newcommand{\dn}{\downarrow}

\section{Introduction}

The pairing interaction is a key concept in nuclear and solid-state
physics, as well as in quantum many-body theory in general. Pairing
between fermions has been studied in bulk materials
\cite{bardeen1957,micnas1990,tinkham}, thin films \cite{meservey1970},
layered materials \cite{cao2018,park2021,nakagawa2021},
atomic nuclei
\cite{Bohr1958,migdal1959,belyaev1961,Richardson1964,zelevinsky2003},
nuclear matter in neutron stars \cite{dean2003}, cold atom systems
\cite{nozieres1985,jochim2003,greiner2003,zwierlein2003,bartenstein2004,ketterle2008},
nanoscopic metal grains
\cite{Averin1992,Tuominen1992,Eiles1993,vonDelft1996,vonDelft2001},
and ultrasmall superconducting islands
\cite{hybrid2010,Krogstrup2015}. When the pairing is weak, the BCS
mean-field approach works well \cite{bardeen1957}. When the pairing is
strong, all fermionic particles pair up into bosons, then the bosons
condense: this produces a BEC. Both regimes are smoothly connected,
despite the fact that the two limits are physically quite different
\cite{nozieres1985,ketterle2008,randeria2014}.

A magnetic impurity, such as a magnetic dopant or a semiconductor
quantum dot (QD), is well known to induce subgap states in the
superconducting gap of a BCS superconductor
\cite{yu1965,shiba1968,rusinov1969,shiba1973,sakurai1970,matsuura1977,yazdani1997,balatsky2006,alloul2009,franke2011,lee2017prb,meden2019review},
which are known as the Yu-Shiba-Rusinov (YSR) states in the limit of well-defined
local moment.
This raises the questions about the persistence of the YSR states all the way to
the BEC limit and about the evolution of their properties in this cross-over. These 
questions are pertinant because the ground state and the elementary
excitations in the two limits have different properties \cite{ketterle2008}.

In this work, we address this subject by coupling a magnetic impurity to a superconductor (SC) described by the Richardson's "picket-fence" pairing model, and tracking the subgap states through the crossover from weak to strong pairing. 
We show that the YSR states not only persist in the cross-over from the BCS to the BEC regime, but do not even change significantly at the quantitative level after an appropriate rescaling of the input parameters and the resulting energies.  
We also establish that a simple model based on the flat-band limit of the Richardson's model (RM) reproduces the full phenomenology of an interacting QD coupled to a SC island.

The RM first appeared as a model of pairing forces in nuclear physics \cite{Richardson1963,Richardson1964,Richardson1966} and was later reintroduced as a description of nanoscale SC grains
\cite{,vonDelft1999,vonDelft2001,Dukelsky2004,Glazman2021}. The SC is modelled as a set of equidistant energy levels representing the time-reversal conjugate single-particle states with all-to-all pairing interaction. 
While the RM reduces to a BCS system in the thermodynamic limit, it has important advantages for smaller systems. Namely, it does not use the mean-field approximation and thus conserves particle number. This makes it a more suitable choice for modelling of mesoscopic SC systems. 
Unlike BCS, it is also applicable for all coupling strengths \cite{Ortiz2005} which is critical in the present work.

The RM is integrable and analytically solvable in terms of hard-core bosons\cite{Richardson1964,Richardson1966}, but coupling it with an impurity level introduces pair-breaking processes and breaks integrability. We have previously found a representation of the QD-SC model as a matrix product operator (MPO), allowing us to use density matrix renormalization group (DMRG) \cite{PhysRevLett.69.2863,SCHOLLWOCK201196,PhysRevB.78.035116} to obtain exact solutions for a QD coupled to one \cite{coulomb1} or two \cite{coulomb3} SC channels. 
With this method we obtain exact properties of a magnetic impurity coupled to an interacting SC with no parameter restrictions, from the BCS to the BEC limits, for all system sizes including ultrasmall SC islands. Good agreement to experiment confirms that this is a suitable description of real systems \cite{coulomb2}. 

This article is organized as follows. In Sec.~\ref{model} we present
the Hamiltonian and the methods that permit its numerical solution. 
In Sec.~\ref{numerical} we present the results for the evolution from
the BCS to the BEC limit. In Sec.~\ref{exact} we
present an exact solution of the model in the limit of a completely
flat band (equivalent to the deep BEC regime). In Sec.~\ref{ZBW} we
discuss how this solution can also be considered as arising from a
ZBW BCS model which leads to exactly the same matrix
representation. We conclude in Sec.~\ref{discuss} with a perspective on the possible
extensions of the flat-band model, e.g. to a two-channel (Josephson junction)
situation. Some lengthy derivations are presented in Appendices \ref{app1}-\ref{norm}.
In the Supplemental material we provide Mathematica notebooks with a computer-algebra verification
of all mathematical statements presented in this work \cite{SM}.
The full source code of the DMRG solver is available on a public repository, including a large set
of examples \cite{rok_zitko_2022_6461063}.
We also provide the input files for the solver and a corresponding Mathematica notebook with
an exact calculation of the ground state energies in the flat-band limit \cite{SM}; this reference
calculation is discussed in Appendix \ref{appD}.

\newcommand{\LL}{\mathcal{L}}

\section{Model and method}
\label{model}

The Richardson's model is \cite{vonDelft2001,coulomb1}
\beq{
H=\sum_{i\sigma} \epsilon_i c^\dag_{i\sigma} c_{i\sigma} 
- G \sum_{i,j} c^\dag_{i\uparrow} c^\dag_{i\downarrow}
c_{j\downarrow} c_{j\uparrow},
}
where $c^\dag_{i\sigma}$ creates a particle in the level $i$ with spin
$\sigma \in \{ \uparrow,\downarrow \}$ and $\epsilon_i$ are the
energy levels spanning the interval $[-D:D]$ spaced by $d=2D/\LL$, where
the half-bandwidth $D \equiv 1$ sets the energy scale, and $\LL$ is the total number of
levels. The indexes $i$
and $j$ range from 1 to $\LL$. The coupling $G$ can also be written as
\beq{
G = \alpha d = D \frac{2\alpha}{\LL} = \frac{g}{\LL},
}
where we have introduced $\alpha$ as the dimensionless strength of the
pairing interaction, and $g$ as the corresponding dimensionfull
strength; $\alpha$ and $g$ are intensive quantities. 

The BCS to BEC limiting process can be implemented in two equivalent
ways: by increasing the coupling $G$ at constant bandwidth $D$, or by
decreasing the bandwidth $D$ at constant coupling $G$.  We opt for the
second possibility in this work. We implement this by appropriately
rescaling the coefficients $\epsilon_i$ in the kinetic-energy term of
the SC Hamiltonian. Going to the BEC limit thus corresponds to taking
the flat-band limit, i.e., omitting the kinetic-energy terms from the
Hamiltonian.

We describe the coupling of the interacting QD to the SC
using an Anderson impurity Hamiltonian
\cite{matthias1958,anderson1961,soda1967,coulomb1}:
\beq{
\begin{split}
\HQD &= \epsQD \hat{n} + \UQD\hat{n}_\uparrow \hat{n}_\downarrow 
+ v \frac{1}{\sqrt{\LL}} \sum_{i,\sigma} \left( c^\dag_{i,\sigma}
d_\sigma + d^\dag_\sigma c_{i,\sigma} \right)  \\
&= \epsQD \hat{n} + \UQD\hat{n}_\uparrow \hat{n}_\downarrow 
+ v \sum_{i,\sigma} \left( f^\dag_\sigma
d_\sigma + d^\dag_\sigma f_{\sigma} \right),
\end{split}
}
where we have defined the bath orbital $f$ through
\beq{
\label{f}
f_\sigma = \frac{1}{\sqrt{\LL}} \sum_{i=1}^\LL c_{i,\sigma}.
}
The QD filling is controlled through the occupancy parameter $\nu$ defined via
\beqz{
\epsQD=\UQD(1/2-\nu).
}
The particle-hole symmetric point of the QD
corresponds to $\epsQD=-U/2$, i.e., $\nu=1$. The hybridisation is
quantified using $\Gamma=\pi \rho v^2$ with the density of states
\beqz{
\rho=1/2D=1/d\LL.
}

We use the DMRG to find the ground state and a few lowest excited states in symmetry sectors with conserved particle number $n$ and z-component of spin $s_z$. We choose $n$ close to half filling with $s_z = 0$ for even $n$ and $s_z = 1/2$ for odd $n$. 

The Hamiltonian is written in MPO form with $9 \times 9$ matrices \cite{coulomb1}. The DMRG results are obtained with system size $\LL = 100$, with maximal MPS matrix dimension of $2000$ and the energy convergence criterion set at $\delta = 10^{-8}$. System sizes up to $\LL = 2000$ with $\delta = 10^{-12}$ are realistically achievable, but increasing the system size produces marginal gains. Using $\LL \gtrsim 30$ already gives good results at much smaller computational expense. 
We set the pairing strength $\alpha = 0.4$ to ensure that an appropriate amount of energy levels participate in pairing even when they span the entire energy range $[-1:1]$. 

The SC-QD models can also be reliably solved by the numerical renormalization group \cite{satori1992,sakai1993,yoshioka1998,yoshioka2000,bauer2007,pillet2013,lee2017prb,supercurrent,Kezilebieke2019,RubioVerdu2021} as well as with continuous-time Quantum Monte Carlo (QMC) using $U$ \cite{Luitz:2010bn} and hybridization \cite{Zonda:2015cd, Pokorny2018, Domanski2017, Zalom2021} expansion. 
These methods require a weakly interacting bath and are thus limited to the mean-field BCS description of superconductivity. 
The RM without the impurity has been treated by QMC in the canonical ensemble \cite{Rombouts2006}, and its strong coupling limit has been investigated by expanding the Richardson's solution in powers of $1/\alpha$ using pseudospin operators \cite{Yuzbashyan2003}. 
Our DMRG approach is currently the only reliable technique to address the full interacting problem in all parameter regimes.

\section{Numerical results for the BCS-BEC cross-over}
\label{numerical}

We introduce $\gamma$ as a parameter that multiplies the kinetic
energy term of the Hamiltonian. The flattening of the band (and the
BCS-BEC crossover) thus correspond to a variation from $\gamma=1$ to
$\gamma=0$. In this section we present the evolution of the subgap
spectra in this crossover.

\subsection{Excitation energies}

\begin{figure*}
\includegraphics[width=16cm]{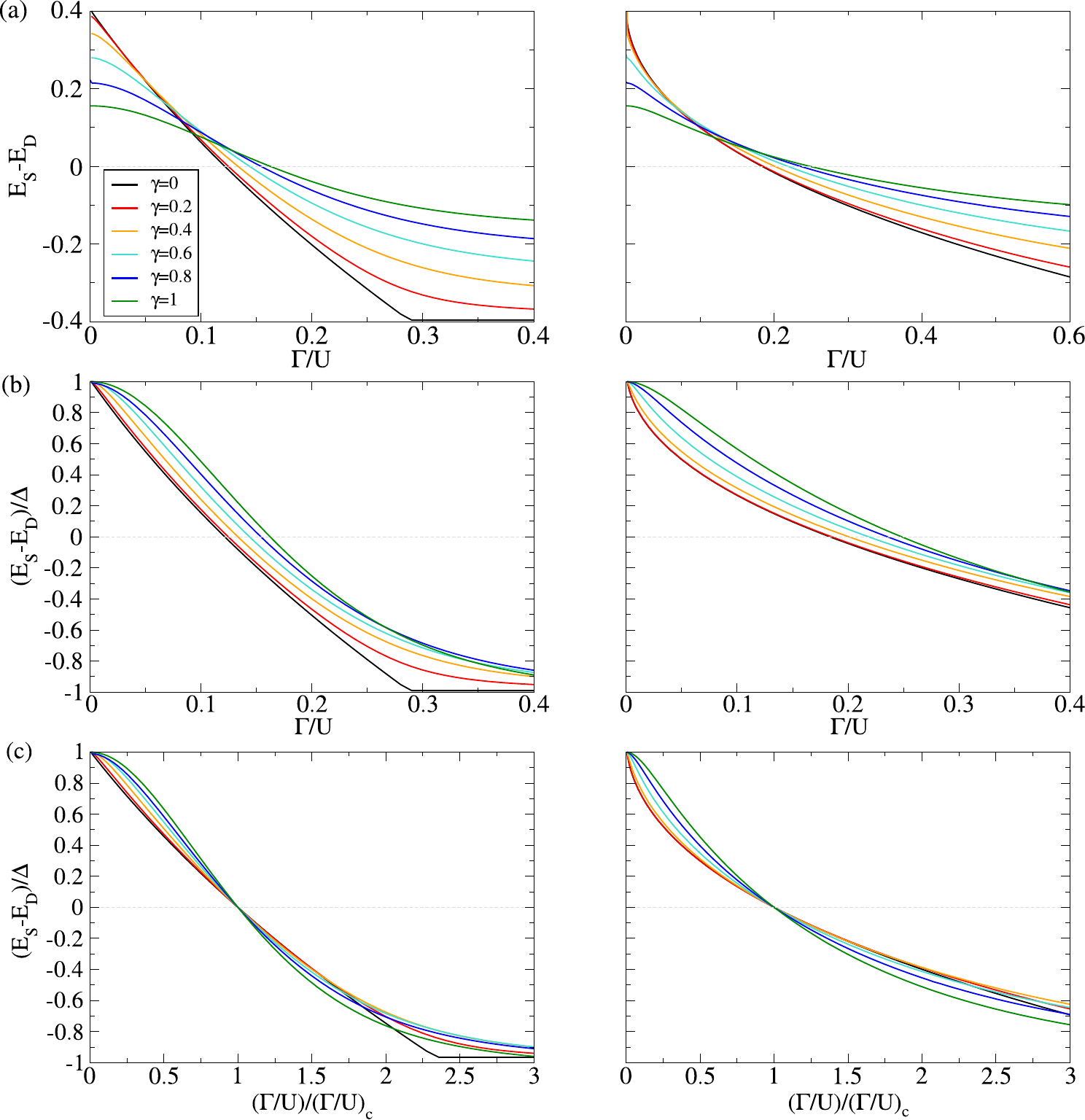}
\caption{Singlet-doublet energy difference $E_\mathrm{YSR}=E_S-E_D$ as
a function of the hybridisation $\Gamma$ for a range of bandwidths
$\gamma$. Energy difference $E_S-E_D$ is expressed either in (a)
absolute units or (b,c) rescaled by the SC gap $\Delta$. 
In (c), the horizontal axis is furthermore rescaled so that the
singlet-doublet transition points coincide. The parameters are
$\alpha=0.4$, strong electron repulsion $U/\Delta_0=40$ (left column)
and moderate repulsion $U/\Delta_0=5$ (right column), where $\Delta_0 \approx 0.16$ is the
gap at $\gamma=1$, and particle-hole symmetric tuning $\nu=1$. The
results have been corrected for the finite-size effects as described
in Ref.~\onlinecite{coulomb1}.}
\label{Gamma}
\end{figure*}

Fig.~\ref{Gamma} shows the excitation energy of the subgap state at
the particle-hole symmetric point ($\epsilon=-U/2$) as a function of
the hybridisation strength $\Gamma$. We consider two cases,
$U/\Delta_0=40$ and $U/\Delta_0=5$, where $\Delta_0$ is the
SC gap in the $\gamma=1$ limit. The first is
representative of the large-$U$ limit where the impurity is a pure
exchange scatterer (Kondo limit) and the subgap states are
well-defined YSR states, while the second is a generic
situation where the singlet subgap states have a mixed character due to a
stronger proximity effect which admixes states of zero and double
impurity occupancy. The first row shows the results without any
rescaling. Withing $\Gamma$ increasing from zero, we see a singlet
state detaching from the continuum of Bogoliubov states: this is 
the YSR singlet. It crosses the zero energy-difference line at $\Gamma=\Gamma_c$ at
which point it becomes the ground state of the system.

We now rescale the results in two steps: 1) we rescale the
energies in terms of the SC gap at given $\gamma$ (second
row), 2) we furthermore rescale the $\Gamma/U$ axis so that the
singlet-doublet transition points for all $\gamma$ coincide (third
row). These fully rescaled results demonstrate a reasonably good overlap,
showing that during the BCS to BEC crossover the changes in the YSR excitation
energy are only quantitative and actually rather small. This is one of the key
results of this work.

\begin{figure}
\includegraphics[width=8cm]{{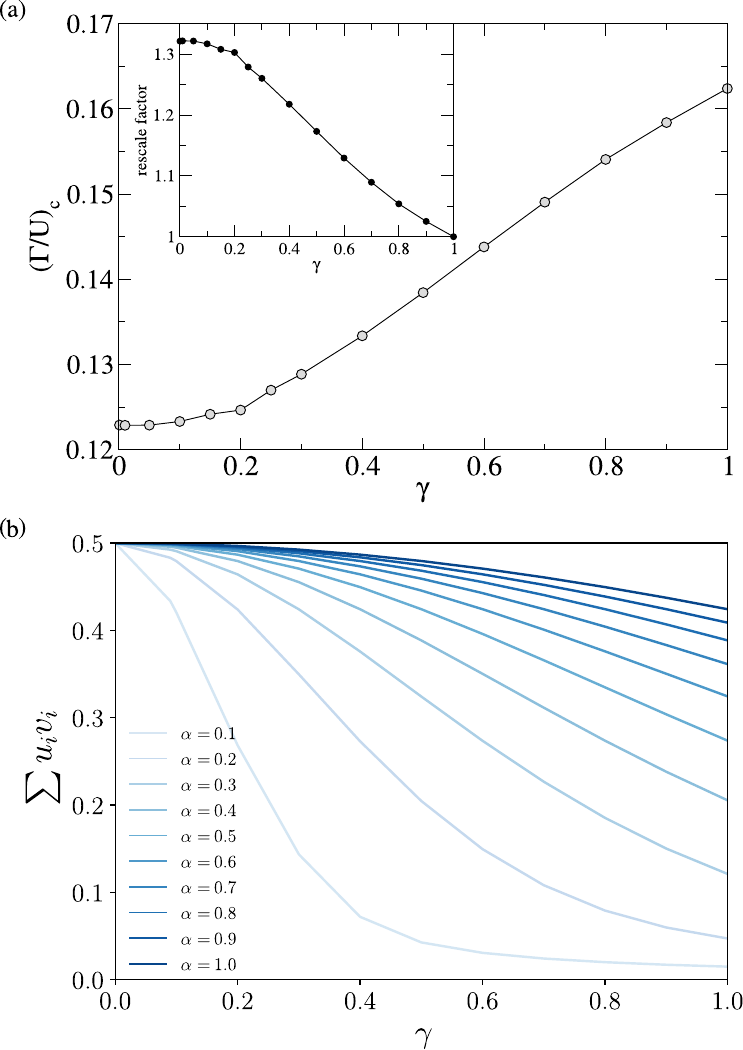}}
\caption{
(a) Bandwidth $\gamma$ dependence of the transition point
$(\Gamma/U)_c$. The
inset shows the rescaling factor for the hybridisation strength
that would be required to obtain equivalent results.
(b) Pairing strength $\bar{\Delta}$ dependence on $\gamma$. It measures the proportion of levels which contribute in pairing, with all levels contributing equally when $\bar{\Delta} = 1/2$.
}
\label{gamma}
\end{figure}

Fig.~\ref{gamma}(a) presents the transition point $(\Gamma / U)_c$ extracted from Fig.~\ref{Gamma}. The hybridization rescaling factor used in the bottom row of Fig.~\ref{Gamma} is shown in the inset. 
By decreasing the bandwith $\gamma$ we allow more levels to participate in hopping processes resulting in larger effect of hybridization $\Gamma$.  The transition point thus moves to smaller value of $\Gamma$ with decreasing bandwidth $\gamma$.

In Fig.~\ref{gamma}(b) we plot the $\gamma$ dependence of the pairing correlations $\sum_i u_i v_i$, where $u_i = \langle c_{i\downarrow} c_{i\uparrow} c^\dag_{i\uparrow} c^\dag_{i\downarrow} \rangle ^{1/2}$ and $v_i = \langle c^\dag_{i\uparrow} c^\dag_{i\downarrow} c_{i\downarrow} c_{i\uparrow} \rangle^{1/2}$, for a range of pairing interaction strengths $\alpha$. The sum measures the proportion of levels that contribute to pairing, with all levels contributing equally when $\sum_i u_i v_i = 1/2$. Similarly as in Fig.~\ref{gamma}(a), the decrease in bandwidth allows more levels to contribute in pairing processes, strengthening the correlations. 
Increasing the pairing strength $\alpha$ at given bandwidth results in qualitatively the same crossover as decreasing the bandwidth $\gamma$, illustrating the two possible ways of probing the BEC - BCS crossover. 
The pairing correlations in the zero bandwidth limit are further discussed in section \ref{pairing}.

\begin{figure}[htbp]
\includegraphics[width=8cm]{{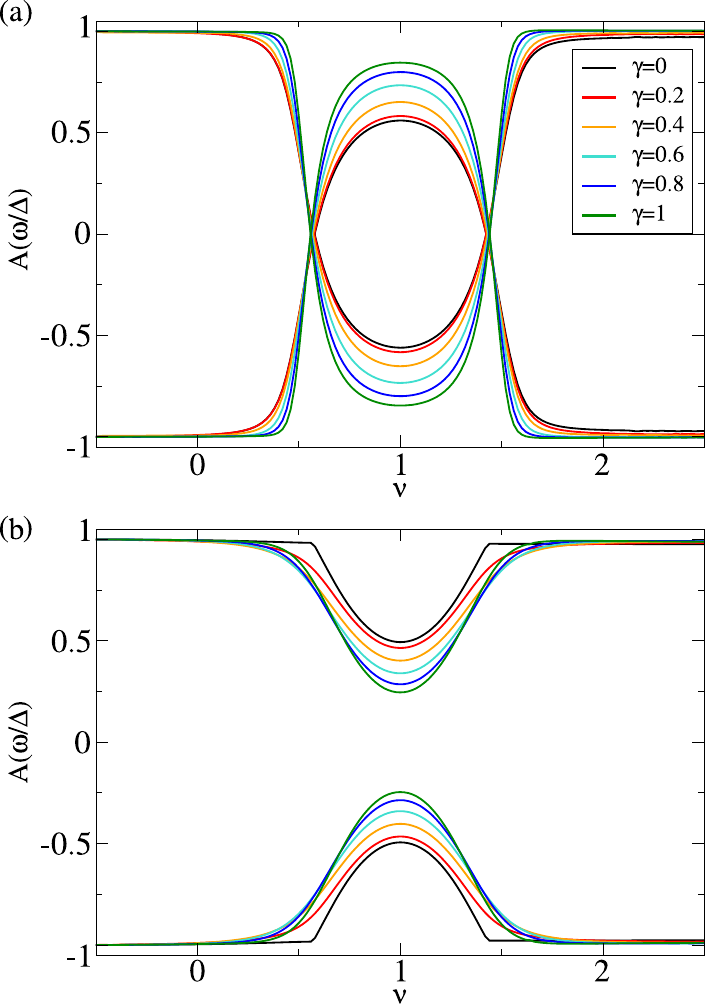}}
\caption{Gate-voltage dispersion of the subgap states for two values of hybridisation: (a) $\Gamma/U=0.05$, (b) $\Gamma/U=0.2$. Other parameters are $\alpha=0.4$, $U/\Delta_0=40$.
The frequencies are rescaled by $\Delta$.}
\label{dispersion2}
\end{figure}

In Fig.~\ref{dispersion2} we show the gate-voltage dispersion for two
fixed values of $\Gamma/U$, on either side of the singlet-doublet
transition value $\Gamma_c$. The energies are scaled by the gap. We
find a good overlap, which could be further improved by adjusting the
$\Gamma$ values as described above.

\subsection{Subgap state properties}
\begin{figure*}[htbp]
\includegraphics[width=16cm]{{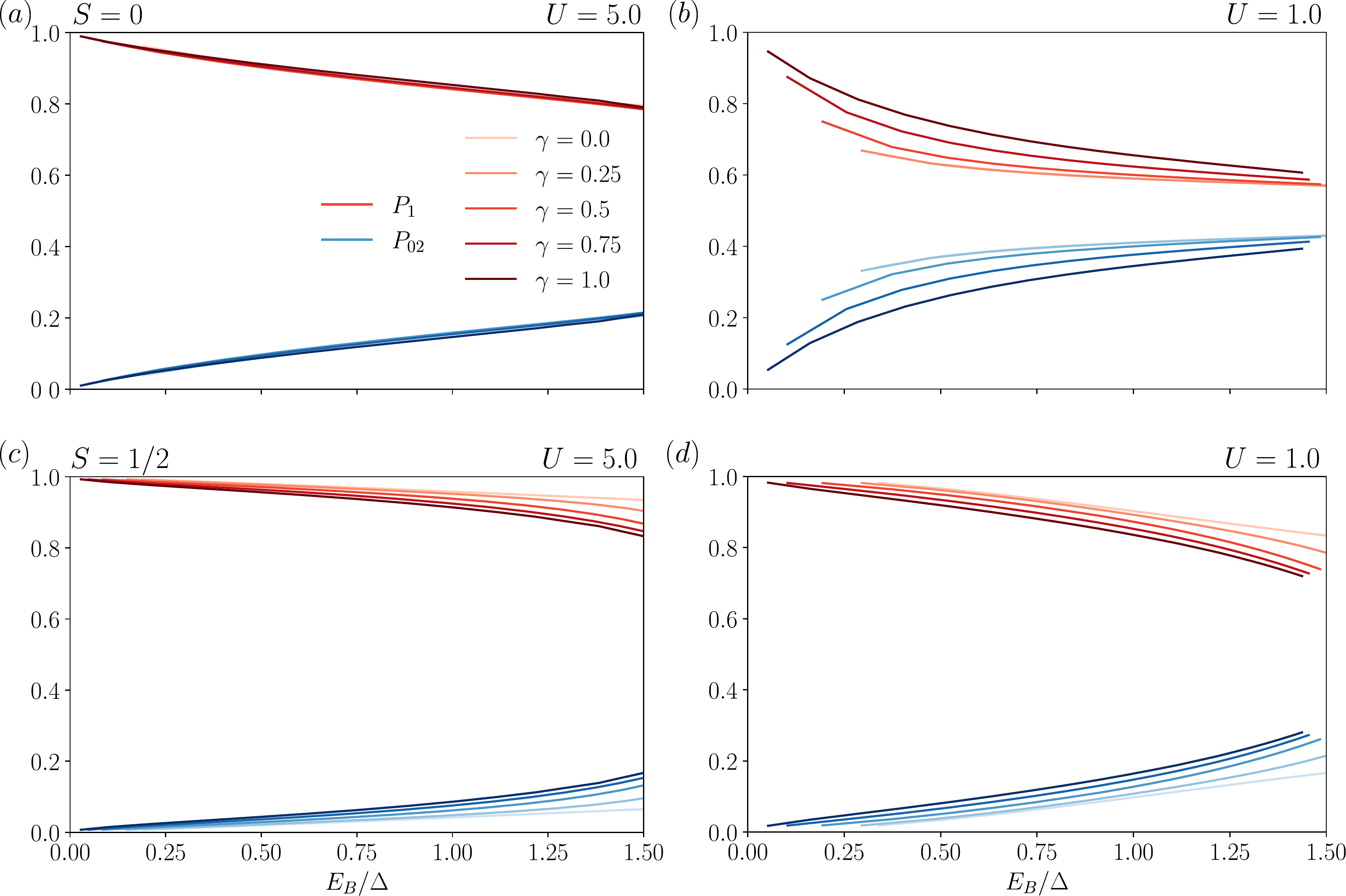}}
\caption{
Impurity occupation probabilities $P_1$ and $P_{02}$ versus binding energy $E_B=E_S-(E_D+\Delta)$ for different values of bandwidth $\gamma$ for the singlet (top) and doublet (bottom) ground state. $E_B$ is the energy gain of the subgap singlet state compared to the first excited (decoupled) singlet state and is rescaled by the optical gap $\Delta$ for each $\gamma$. The pairing interaction strength is $\alpha=0.4$.
}
\label{Pimp}
\end{figure*}

In addition to the approximate overlap of the YSR state energies, we also find a high degree of similarity in the state properties across the crossover. As an example, in Fig.~\ref{Pimp} we plot the probabilities that the impurity level is singly occupied ($P_1$) or doubly occupied/empty ($P_{02}$) versus binding energy $E_B=E_S-(E_D+\Delta)$. These are obtained from the reduced density matrix for the impurity site and contain important information about the nature of the states.

In the top row we plot the singlet ground state. For the YSR state in a superconductor we expect $P_\uparrow = P_\downarrow = 1/2$ and thus $P_1 = 1$. With increasing $E_B$ (by increasing the hybridization $\Gamma$) the local magnetic moment of the impurity level decreases due to larger charge fluctuations. $P_{02}$ measures the contribution of the proximitized ABS state, which becomes dominant in the limit of large $\Gamma/U$. Fig.~\ref{Pimp}(a) shows that this behaviour persists throughout the BCS-BEC transition. In this case with $U \gg \Delta$ we find good overlap of the curves. The situation away from this limit, where $U \approx \Delta$, is shown in Fig.~\ref{Pimp}(b). Here the magnitude of charge fluctuations depend not only on $\Gamma/U$ but also on $\Gamma/\Delta$. The fact that two different scales control the charge fluctuations decreases the quality of the collapse, but the qualitative agreement for different $\gamma$ remains.

The bottom row shows the doublet ground state, where the mechanism of changing state character is a bit different. Increasing $E_B$ increases the admixture of the excited doublet state, which is identical to the YSR singlet with an additional free quasiparticle in the bath. This excited state transforms from a YSR-like singlet to an ABS-like proximitized state (as discussed in the previous paragraph), resulting in the decrease of $P_1$, which is, however, more gradual compared to the singlets. This is most obvious when comparing the two cases with smaller $U$ in Fig.~\ref{Pimp}(b) and Fig.~\ref{Pimp}(d).

We note that we find a similar qualitative agreement for different $\gamma$ for many state properties. This is an important result in its own, as it implies that it is possible to use the flat-band solutions to investigate the nature of the low energy states in realistic QD--SC systems.

\section{Exact solution for the flat-band limit}
\label{exact}

In the flat-band limit ($\gamma=0$, i.e., $D=0$ or $\epsilon_i \equiv 0$), two
renormalisation processes associated with the kinetic energy of
band electrons no longer occur. These are the renormalization of the
exchange interaction from the bare Kondo exchange coupling
$J_K=8\Gamma/\pi \rho U$ to the scale of the exponentially lower Kondo
temperature $T_K=\exp(-1/\rho J_K)$, and the renormalization of the pairing
interaction from the bare scale of attractive coupling $g$ to the scale of the exponentially lower SC gap
$\Delta=\exp(-1/\rho g)$. Both renormalization processes have exactly
the same origin: the kinetic energy (dispersion) of the conduction band electrons.
In the following we show that the flat-band limit is analytically
solvable, in the sense of being reducible to numerical
diagonalisations of very small matrices. This model provides a
conceptually clear picture of all phenomena occurring in ultrasmall
SCs and provides exact benchmark results for testing other approaches.

\subsection{Eigenstates of the superconductor}

We take the $\epsilon_i \equiv 0$ limit of the RM
and write the SC part of the Hamiltonian as
\beq{
H
= -\frac{g}{\LL} \sum_{i,j} c^\dag_{i\uparrow} c^\dag_{i\downarrow}
c_{j\downarrow} c_{j\uparrow} 
= -\frac{g}{\LL} \sum_{i,j} A^\dag_i A_j,
}
where $A_j = c_{j\downarrow} c_{j\uparrow}$ are the
hard-core boson operators with the commutation rule:
\beq{
[A_i,A_j^\dag]=\delta_{ij}(1-\hat{n}_i),
\label{com1}
}
where $\hat{n}_i = \sum_\sigma c^\dag_{i\sigma}c_{i\sigma}$ is the level occupancy operator.
The following relations also hold:
\beq{
\begin{split}
A_i^2 &\equiv 0, \\
(A_i^\dag)^2 &\equiv 0, \\
A_i^\dag A_i &= \hat{P}_{2,i} = \hat{n}_{i\uparrow} \hat{n}_{i\downarrow},\\
A_i A_i^\dag &= \hat{P}_{0,i} = 1-\hat{n}_i + \hat{n}_{i\uparrow} \hat{n}_{i\downarrow}.
\end{split}
}
The operators $\hat{P}_{0,i}$ and $\hat{P}_{2,i}$ are the projectors on the subspace
with occupancy 0 or 2 on level $i$. 

The Hamiltonian $H$ is not particle-hole (p-h) symmetric. The symmetry
can be restored by adding a potential energy term:
\beq{
\tilde{H} = H + \tilde{\epsilon} \sum_i \hat{n}_i,
}
with
\beqz{
\tilde{\epsilon}=\frac{g}{2\LL} = \frac{\alpha d}{2} = \frac{\alpha}{\LL},
}
using $D
\equiv 1$ as energy unit.

\newcommand{\UU}{\mathcal{U}}
\newcommand{\BB}{\mathcal{B}}

Let $\UU$ be the set of ``unblocked levels'' (those occupied by either
0 or 2 electrons). The blocked levels $\BB$ (those with occupancy 1) do not
participate in pairing and they fully decouple from the problem in the
absence of an impurity, thus in this subsection all the sums that
follow are constrained to the set of unblocked levels. We will also
use the same symbol $\UU$ to denote the number of unblocked levels,
since this leads to no ambiguity. In the absence of quasiparticles no
level is blocked, $\BB=\emptyset$, and we have $\UU=\LL$.

We introduce the equal-weight linear combination of hard-core
operators corresponding to all unblocked levels,
\beq{
B= \frac{1}{\sqrt{\UU}} \sum_i^\UU A_i,
}
so that
\beq{
H = -g \frac{\UU}{\LL} B^\dag B + H_\mathrm{blocked}.
}
Dropping the blocked subspace from this point on, the full Hamiltonian including the p-h symmetry 
restoring energy shift is thus
\beq{
\label{eq11}
\tilde{H} = -g \frac{\UU}{\LL} B^\dag B + \frac{\alpha}{\LL} \hat{N},
}
where 
\beq{
\hat{N}=\sum_i^\UU \hat{n}_{i}
}
is the total number of electrons in the {\it unblocked} subset of levels (i.e., the energy of unpaired electrons in the 
blocked levels is not included).

\newcommand{\NN}{\mathcal{N}}

The eigenstates for $M$ Cooper pairs in the $\UU$ levels are
\beq{
\Psi^{\LL,\UU}_M = \NN_M |M\rangle = \NN_M (B^\dag)^M |0\rangle,
}
where $\ket{M} = (B^\dag)^M \ket{0}$ and $\NN_M$ is a normalization
factor (see Appendix~\ref{app3}). In the following, the superscripts
$\LL,\UU$ in $\Psi^{\LL,\UU}_M$ will be dropped when no ambiguity may result;
when a single superscript is indicated, it corresponds to $\UU$.

We have $H\Psi_0 = 0$ and $H \Psi_1 = - g(\UU/\LL) \Psi_1$.
General eigenvalues can be computed by recursion (see Appendix~\ref{app1}), 
finding
\beq{
H \ket{M} = -g (\UU/\LL) c_M \ket{M}
}
with
\beq{
c_M = \frac{(1+\UU-M)M}{\UU}.
}
The ground state energy of a system of size $\LL$ with $\UU$ unblocked levels containing 
$M$ Cooper pairs is thus
\beq{
E^{\LL,\UU}_M 
= - 2\alpha \frac{(1+\UU-M)M}{\LL},
}
while the shifted eigenvalues (corresponding to $\tilde{H}$) are
\beq{
\tilde{E}^{\LL,\UU}_M = - 2\alpha \frac{(\UU-M)M}{\LL}.
}
The interpretation of this expression is very simple. Each of the $M$
Cooper pairs resonates with the $\UU-M$ empty levels in the system,
each combination contributing $2\alpha/\LL=G$ to the total energy.
We note that $\tilde{E}$ does not include the energy of single electrons
in the blocked levels, which is not included in Eq.~\eqref{eq11}. We therefore
define the total energy $\hat{E}$ as
\beqz{
\begin{split}
\hat{E}^{\LL,\UU}_M &= \tilde{E}^{\LL,\UU}_M + \frac{\alpha}{\LL}(\LL-\UU) \\
&= \alpha \frac{-2(\UU-M)M+\LL-\UU}{\LL}.
\end{split}
}

\newcommand{\EE}{\mathcal{E}}

For half filling, $n=\LL$ and $M=\LL/2$, and one has for $\UU=\LL$
\beq{
\label{EE}
\tilde{E}^{\LL,\LL}_{\LL/2} = -\alpha \frac{\LL}{2} \equiv \EE.
}
Also, we find
\beq{
\tilde{E}^{\LL,\LL}_{\LL/2+1} = \EE + \frac{2\alpha}{\LL},\,\,
\tilde{E}^{\LL,\LL}_{\LL/2+2} = \EE + \frac{8\alpha}{\LL},\ldots
}
thus additional Cooper pairs have an energy cost of order $1/\LL$. In
the thermodynamic limit ($\LL\to\infty$), the ground state would be
macroscopically degenerate, as is the case in bulk SCs.

We now consider a single quasiparticle in the system, i.e., we make
one of the $\LL$ levels singly occupied, so that $\UU=\LL-1$. The
energy cost in the $n=\LL+1$ charge sector is
\beq{
E^{qp,+} = \left( \tilde{E}^{\LL,\LL-1}_{\LL/2} + \tilde{\epsilon} \right) - \mathcal{E} =
\left(1+\frac{1}{\LL}\right) \alpha.
}
and in the $n=\LL-1$ charge sector it is
\beq{
E^{qp,-} = \left( \tilde{E}^{\LL,\LL-1}_{\LL/2-1} + \tilde{\epsilon} \right) - \mathcal{E} =
\left(1+\frac{1}{\LL}\right) \alpha.
}
In the equal-charge ($n=\LL$) sector, the lowest excitation contains two quasiparticles. Its energy is
\beq{
\left( \tilde{E}^{\LL,\LL-2}_{\LL/2-1} + 2\tilde{\epsilon} \right) -
\mathcal{E} = 2\alpha.
}
The cost of a quasiparticle excitation in flat-band superconductors is equal to the bare
pairing interaction \cite{belyaev1961,khodel1990,volovik2018}.

\subsubsection*{Pairing strength}
\label{pairing}
The strength of pairing correlations may be quantified by evaluating the following correlators \cite{Braun1999}:
\beq{
\begin{split}
\bar{v}_i &= \langle c^\dag_{i\uparrow} c^\dag_{i\downarrow} c_{i\downarrow} c_{i\uparrow} \rangle
^{1/2} = \langle A_i^\dag A_i \rangle^{1/2}, \\
\bar{u}_i &= \langle c_{i\downarrow} c_{i\uparrow} c^\dag_{i\uparrow} c^\dag_{i\downarrow} \rangle
^{1/2}  = \langle A_i A_i^\dag \rangle^{1/2}, \\
\bar{\Delta} &= \frac{g}{\LL} \sum_i \bar{v}_i \bar{u}_i, \\
\bar{\Delta}' &= \frac{g}{\LL} \sum_i 
\Bigl( 
\langle n_{i\uparrow} n_{i\downarrow} \rangle - 
\langle n_{i\uparrow} \rangle 
\langle n_{i\downarrow} \rangle
\Bigr)^{1/2}.
\label{eq:SC_pairing}
\end{split}
}
$\bar{v}_i$ is the amplitude to find a pair of states occupied, while
$\bar{u}_i$ is the amplitude to find a pair of states empty. If the
system is homogeneous, all amplitudes are the same, $\bar{v} \equiv
\bar{v}_i$ and $\bar{u} \equiv \bar{u}_i$, so that
\beq{
\bar{\Delta} = g \bar{v} \bar{u}.
}
Both $\bar{\Delta}$ and $\bar{\Delta}'$ reduce to
$\Delta_\mathrm{BCS}$ in the thermodynamic limit \cite{vonDelft1999}.

At half filling, the results is obviously $\bar{v}^2=\bar{u}^2=1/2$, because there is probability
$1/2$ to find any given level unoccupied and the same probabilty to find it doubly occupied, and
by construction it cannot be singly occupied, since no level is blocked.
Thus
\beq{
\bar{\Delta} = \frac{g}{2} = \alpha.
}
In general, the amplitudes $\hat{v}_i$ and $\hat{u}_i$ are related,
since $A_i^\dag A_i = A_i^\dag A_i - (1-\hat{n}_i)$, thus
\beq{
\bar{v}^2 = \bar{u}^2 - \bra{\Psi_M} (1-\hat{n}_i) \ket{\Psi_M} 
= \bar{u}^2 - 1 + \frac{2M}{\UU}.
}
In the subspace where all levels are either unoccupied or doubly occupied, we also have
$\bar{v}^2+\bar{u}^2=1$. Then it follows immediately that
\beq{
\bar{v}^2 = \frac{M}{\UU},
\quad
\bar{u}^2 = \frac{\UU-M}{\UU},
\quad
\bar{\Delta} = 2\alpha \frac{\sqrt{M(\UU-M)}}{\UU}.
}

The other estimate, $\bar{\Delta}'$, can also be easily computed:
\beq{
\bar{\Delta}' = \frac{g}{\LL} \UU \left( \frac{M}{\UU} - \frac{M}{\UU} \frac{M}{\UU} \right)^{1/2},
}
because in a homogeneous system $\langle \hat{n}_{i\uparrow} \hat{n}_{i\downarrow} \rangle = \langle \hat{P}_{2i} \rangle = M/\UU$
and $\langle n_{i\sigma} \rangle = M/\UU$. At half filling, this yields 
\beqz{
\bar{\Delta}' = \alpha.
}

In this flat-band model, the pairing correlations are thus
directly proportional to the coupling-strength $\alpha$. This is a
known feature of strong-coupling SC systems
\cite{belyaev1961,khodel1990,volovik2018}. The exponential reduction
in the standard weak-coupling BCS theory is a consequence of the
finite width of the electron band. 

\subsection{YSR states for a Kondo impurity}
\label{kondo}

We now include a magnetic impurity into consideration. First we
consider the case of a pure exchange scatterer and take the large-$U$
limit of the impurity Hamiltonian at $\nu=1$ to obtain an effective
Kondo Hamiltonian \cite{schrieffer1966,bravyi2011}:
\beq{
\HQD = \frac{J}{\LL} \sum_{i,j} \mathbf{S} \cdot \mathbf{s}_{i,j}, 
}
where the Kondo exchange coupling is
\beq{
\label{eq29}
J = 8v^2/U=\frac{16}{\pi} (\Gamma/U),
}
and the inter-level spin operators are defined as 
\beq{
\mathbf{s}_{i,j} = \frac{1}{2} c^\dag_{i,\alpha}
\boldsymbol{\sigma}_{\alpha,\beta} c_{j,\beta}.
}
Here $\boldsymbol{\sigma}=\{ \sigma^x,\sigma^y,\sigma^z \}$ is a
vector of Pauli matrices and the expression is summed over repeated
spin indexes $\alpha$ and $\beta$. Note that in this discrete model
$\Gamma$ does not have the significance of a level width, but it
nevertheless quantifies the coupling strength. Equivalently, we may
write
\beq{
\HQD = J \mathbf{S} \cdot \mathbf{s},
}
with $\mathbf{s}$ the spin of an electron in the orbital $f$, see
Eq.~\eqref{f},
\beq{
\mathbf{s} = \frac{1}{2} f^\dag_{\alpha}
\boldsymbol{\sigma}_{\alpha,\beta} f_{\beta}.
}
For half filling with $M=\LL/2$ Cooper pairs (assuming even $\LL$) and one
electron in the impurity level (odd total number of electrons $n=\LL+1$,
i.e., spin-doublet sector), the impurity remains entirely decoupled
from the SC in the ground state. The reason is that the
product state $\psi = \ket{M} \otimes \ket{\sigma}$ is an eigenstate
of $\HQD$. First we note that the state $\ket{M}$ is an eigenstate of
$\hat{s}_z$ with $\hat{s}_z \ket{M}=0$. We thus only need to consider
the transverse $S^- s^+$ terms. This operator is composed of pairs
\beq{
c^\dag_{i\uparrow} c_{j\downarrow} + c^\dag_{j\uparrow} c_{i\downarrow}.
}
We have
\beq{
[c^\dag_{i\uparrow} c_{j\downarrow}, A_j^\dag] =
-c^\dag_{i\uparrow} c^\dag_{j\uparrow},
\quad
[c^\dag_{j\uparrow} c_{i\downarrow}, A_i^\dag] = 
c^\dag_{i\uparrow} c^\dag_{j\uparrow}.
}
This leads to a cancellation for any $i\neq j$ pair. Furthermore, the diagonal $i=j$ terms are
trivially zero due to Pauli exclusion principle. Thus we have $s^+ \ket{M}=0$, showing that
$\ket{M}$ is indeed an eigenstate.

The same commutation relations also establish that the pure exchange
interaction preserves the decoupling between the blocked (singly
occupied) levels $\BB$ and the unblocked (zero or double occupancy)
levels $\UU$. Thus the Hamiltonian in the subspace of $\BB$ blocked levels is
\beq{
H_\BB = J \mathbf{S} \cdot \mathbf{s},
}
where $\mathbf{s}$ is now constrained only over the blocked levels.
This is an exact statement. The ground state in the even occupancy
$n=\LL+2$ sector results from the interaction between the single
quasiparticle and the impurity spin (i.e., the YSR singlet state).
This is clearly a simple two-body problem of two spins coupled by an
isotropic exchange interaction, with one singlet eigenstate at energy
$-3/4J$ and one triplet eigenstate at energy $+1/4J$. The creation of
the quasiparticle has an energy cost of $(1+1/\LL)\alpha$. The
excitation energy is thus
\beq{
E_\mathrm{YSR} = \left( 1+ \frac{1}{\LL} \right) \alpha - \frac{3J}{4}.
}
The YSR singlet binding energy is
\beq{
\Delta E = E^{qp,+}-E_\mathrm{YSR}=\frac{3}{4} J = \frac{12}{\pi} (\Gamma/U).
}
There is a YSR triplet state in the ``continuum'' with an energy surplus of $1/4J$.

\subsection{YSR states for an Anderson impurity}

At finite $U$, an electron from a doubly occupied bath level can hop on the impurity site, and an
electron can hop from the impurity site to an unoccupied bath level. The decoupling between the
blocked levels $\BB$ and the unblocked levels $\UU$ is lost: the subsectors mix. We need to separately discus
the cases with different fermionic parity (and consequently total spin).

\newcommand{\Stot}{S^\mathrm{tot}}

\subsubsection{Odd total number of electrons (spin-doublet)}

We first consider the case with an odd total number of electrons. The
$z$ component of total spin is a conserved quantity; we will work in
the $s_z=1/2$ subspace. The following normalized states are coupled by
the hopping term:
\beq{
\label{eq79}
\begin{split}
\psi_{0,b} 
&= \ket{0} \otimes c^\dag_{b \uparrow} \Psi^{\LL\backslash b}_M, \\
\Psi_{1} 
&= \ket{1} \otimes \Psi^{\LL}_M, \\
\psi_{1,bb'}
&= \ket{1} \otimes c^\dag_{b\uparrow} c^\dag_{b'\downarrow} \Psi^{\LL \backslash b,b'}_{M-1}, \\
\psi_{2,b} 
&= \ket{2} \otimes c^\dag_{b\uparrow} \Psi^{\LL \backslash b}_{M-1}.
\end{split}
}
The ket $\ket{i}$ represents the impurity state with $i$ electrons,
defined as $\ket{1}=d^\dag_\uparrow \ket{0}$ and $\ket{2}=d^\dag_\uparrow d^\dag_\downarrow \ket{1}$.
In $\psi_{0,b}$ and $\psi_{2,b}$ the level $b$ is blocked; here $1 \leq b
\leq \LL$. The notation $\LL \backslash b$ in the superscript denotes
that the set of unblocked levels $\UU$ encompasses all levels except $b$,
while the subscript represent the number of Cooper pairs. 
In a similar vein, in $\psi_{1,bb'}$ two levels are blocked.
We note that states with two quasiparticles with equal spin in the superconductor (and opposite
spin in the QD, such that the total spin is $S=1/2$ and $S_z=1/2$)
do not couple with the subset of states in Eq.~\eqref{eq79} in the flatband limit.

We now form normalized equal-weight superpositions of $\psi_{0,b}$ and $\psi_{2,b}$,
\beqz{
\label{eq81}
\Psi_{0} = \frac{1}{\sqrt{\LL}} \sum_{b=1}^\LL \psi_{0,b}, \quad
\Psi_{2} = \frac{1}{\sqrt{\LL}} \sum_{b=1}^\LL \psi_{2,b},
}
as well as
\beqz{
\begin{split}
\Psi'_1 &= \frac{1}{\sqrt{\LL(\LL-1)}} \sum_{b\neq b'} \psi_{1,bb'}.
\end{split}
}
Note that the double sum is over all $\LL$ so the two quasiparticles form a singlet state.
For large $\LL$ this simplifies to
\beqz{
\Psi'_1 \approx \frac{1}{\LL} \sum_{b\neq b'} \psi_{1,bb'}.
}

The energies of the doublet states $\Psi$ are
\beq{
\begin{split}
E^D_{0} &= {\tilde E}^{\LL,\LL-1}_M + \tilde{\epsilon} - \epsQD, \\
E^D_{1} &= {\tilde E}^{\LL,\LL}_M, \\
E^{D'}_1 &= {\tilde E}^{\LL,\LL-2}_{M-1} + 2\tilde{\epsilon}, \\
E^D_{2} &= {\tilde E}^{\LL,\LL-1}_{M-1} + \tilde{\epsilon} + \epsQD + U.
\end{split}
}
Here we included the energy shifts $\tilde{\epsilon}$ (hence we use energies
$\tilde E$), which are important for this consideration, and we have
subtracted $\epsQD$. The superscript $D$ denotes that these are 
energies in the spin-doublet subspace.

To compute the matrix elements, we make use of an expression derived
in Appendix~\ref{app3} after setting $M=\LL/2$:
\beq{
\label{sqrtfactor}
f^\dag_\sigma \Psi^\LL_{\LL/2} = \frac{1}{\sqrt{2}}
\frac{1}{\sqrt{\LL}} \sum_b
c^\dag_{b,\sigma} \Psi^{\LL\backslash b}_{\LL/2},
}
thus $f^\dag_\sigma d_\sigma \Psi_1 = (1/\sqrt{2}) \Psi_0$.
An analogous expression can be obtained for $f_\sigma
\Psi^\LL_{\LL/2}$,
giving
$d^\dag_\sigma f_\sigma \Psi_1 = (1/\sqrt{2}) \Psi_2$.
In a similar way we obtain all other matrix elements.

\begin{figure}
    \centering
    \includegraphics[width=7cm]{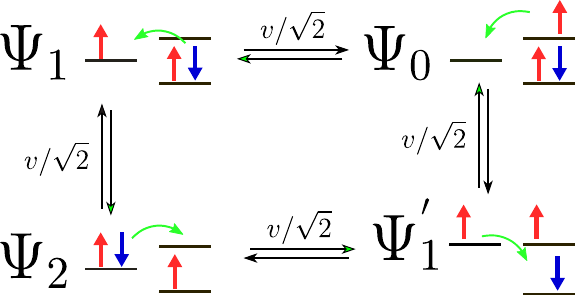}
    \caption{Schematic representation of the states coupled
    in the doublet subspace. In the flat-band limit, no other
    states couple with this subset. Left line: QD level. Right lines: SC levels. Quasiparticles (singly occupied SC levels) represent superpositions of states where the
    single electron occupies any of the superconductor $\LL$ levels.
    }
    \label{fig:D}
\end{figure}

\begin{figure}
\centering
\includegraphics[width=\columnwidth]{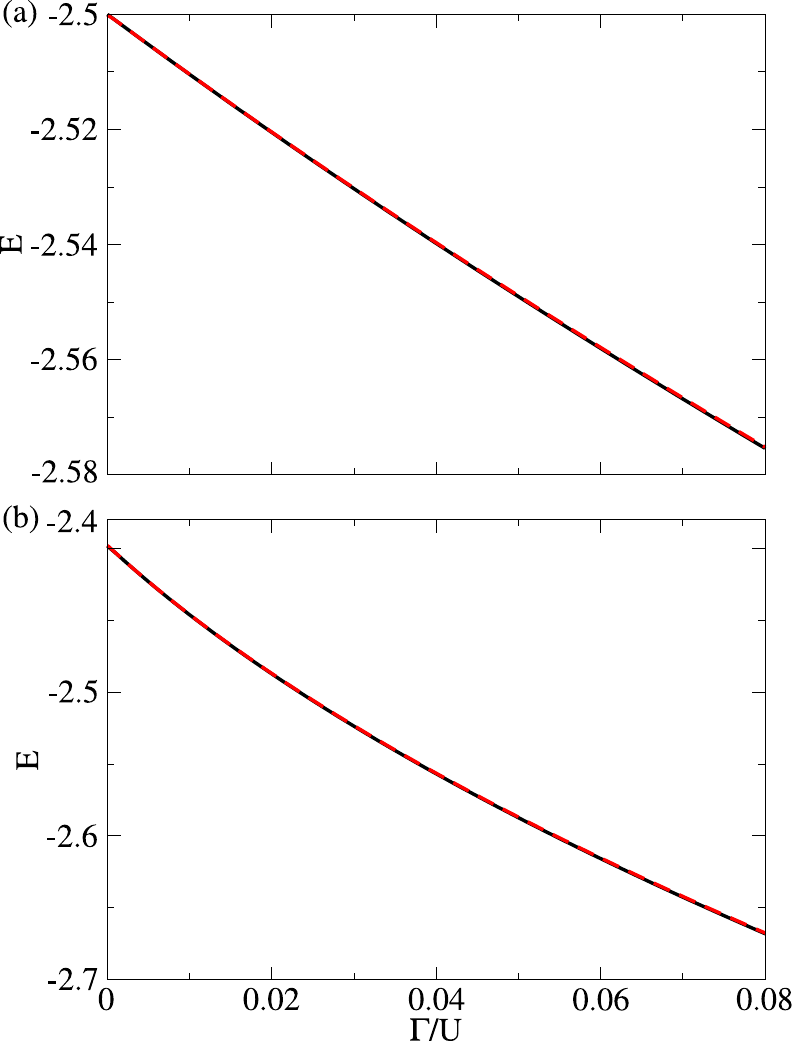}
\caption{Ground state eigenenergies in the flat-band limit for
$U=10\alpha$, specifically $\alpha=0.1$ and $U=1$ in units of $D$,
$\LL=50$. Full line in both panels: DMRG results. (top)
Odd-parity (spin-doublet) sector with $n=\LL+1$, $s_z=1/2$. Dashed
line: lowest eigenvalue of the Hamiltonian in Eq.~\eqref{HD}. 
(bottom) Even-parity (spin-singlet) sector with
$n=\LL+2$, $s_z=0$. Dashed line: lowest eigenvalue of the
Hamiltonian in Eq.~\eqref{H2}. }
\label{figU}
\end{figure}

We obtain a four-level problem with the Hamiltonian in the
orthonormal basis $\{ \Psi_1, -\Psi_1', \Psi_0, \Psi_2 \}$
that can be expressed in the large-$\LL$ limit and at half-filling as
\beq{
\label{HDeff}
H^D_\mathrm{eff} = \begin{pmatrix}
E^D_1 & 0 & v/\sqrt{2} & -v/\sqrt{2} \\
0 & E^{D'}_1 & v/\sqrt{2} & v/\sqrt{2} \\
v/\sqrt{2} & v/\sqrt{2} & E^D_0 & 0 \\
-v/\sqrt{2} & v/\sqrt{2} & 0   & E^D_2 
\end{pmatrix}.
}
The schematic representation of this Hamiltonian is shown in Fig.~\ref{fig:D}.
The full expression for $H^D_\mathrm{eff}$ for arbitrary $\LL$ and $M$ with 
exact matrix elements is given in Appendix~\ref{appD}.
Close to half filling and for small to moderate $v$, the state $\Psi_1'$
is a highly excited states and plays little role (it admixes in the
ground state wavefunction only as a $v^4$ correction). We may thus drop
the second row and second column in the matrix in Eq.~\eqref{HDeff} and write
\beq{
\label{eq82}
H^D_\mathrm{eff} = \begin{pmatrix}
E^D_1 & v/\sqrt{2} & -v/\sqrt{2} \\
v/\sqrt{2} &  E^D_0 & 0 \\
-v/\sqrt{2} &  0   & E^D_2 
\end{pmatrix}.
}
The eigenenergies can be obtained from the cubic roots. At half filling in the particle-hole symmetric point ($\epsQD=-U/2$) the problem simplifies 
even further. We then have
\beq{
\label{eqED}
\begin{split}
E^D_0 = E^D_2 &= -\alpha\frac{\LL}{2} + \alpha + U/2 \equiv E^D_{02}, \\
E^D_1 &= -\alpha\frac{\LL}{2}.
\end{split}
}
The eigenvector $(0,1,1)$ with eigenvalue $E^D_{02}$ decouples.
The ground state can then be obtained by diagonalising the $2 \times 2$
matrix
\beq{
\label{HD}
H^D_\mathrm{eff} = \begin{pmatrix}
E^D_1 & v \\
v & E^D_{02} \\
\end{pmatrix}.
}
Reference calculations demonstrating that the method is exact are presented in Appendix~\ref{appD}.
Even using the approximate expressions from this section which are strictly correct only in the $\LL\to\infty$ limit,
we find excellent agreement with the DMRG results even at moderate $\LL=50$, see Fig.~\ref{figU} (top panel) 
where we plot the lowest eigenvalue of $H^D_\mathrm{eff}$ using the dashed line and the reference
results as full line.

\subsubsection{Even total number of electrons (spin-singlet)}

We now consider the case with an even total number of electrons,
specifically $n=2M+2$, in the spin singlet state. We first enumerate
all relevant states in the subspace with $s_z=0$. There are two states
with single occupancy of the impurity, while in the sectors with
$\nimp=0$ and $\nimp=2$ we need to include two types of states, those
with no quasiparticles and those with two quasiparticles:
\beq{
\begin{split}
\Psi_{0}  &= \ket{0} \otimes \Psi^\LL_{M+1}, \\
\psi'_{0,bb'} &= \ket{0} \otimes c^\dag_{b\up} c^\dag_{b'\dn}
\Psi^{\LL\backslash b,b'}_{M}, \\
\psi_{1,b,\sigma} &= \ket{\sigma} \otimes c^\dag_{b\bar{\sigma}}
\Psi^{\LL\backslash b}_{M}, \\
\Psi_{2}  &= \ket{2} \otimes \Psi^\LL_{M}, \\
\psi'_{2,bb'} &= \ket{2} \otimes c^\dag_{b\up} c^\dag_{b'\dn}
\Psi^{\LL\backslash b,b'}_{M-1},
\end{split}
}
with energies
\beq{
\begin{split}
E^S_{0}  &= {\tilde E}^{\LL,\LL}_{M+1} - \epsQD, \\
{E^S_0}' &= {\tilde E}^{\LL,\LL-2}_{M}+2\tilde{\epsilon}-\epsQD, \\
E^S_{1}  &= {\tilde E}^{\LL,\LL-1}_M + \tilde{\epsilon}, \\
E^S_{2}  &= {\tilde E}^{\LL,\LL}_{M} + \epsQD + U, \\
{E^S_2}' &= {\tilde E}^{\LL,\LL-2}_{M-1}+2\tilde{\epsilon}+\epsQD+U.
\end{split}
}
For one electron beyond half filling (i.e., for $M=\LL/2$) and $\epsQD=-U/2$,
this yields
\beq{
\begin{split}
E^S_0    &= -\alpha\frac{\LL}{2} + U/2 + \frac{2\alpha}{\LL},\\
{E^S_0}' &= -\alpha\frac{\LL}{2} + 2\alpha + U/2 + \frac{2\alpha}{\LL}
= E^S_0 + 2\alpha, \\
E^S_1    &= -\alpha\frac{\LL}{2} + \alpha + \frac{\alpha}{\LL}, \\
E^S_2    &= -\alpha\frac{\LL}{2} + U/2, \\
{E^S_2}' &= -\alpha\frac{\LL}{2} + 2\alpha + U/2 = E^S_2 + 2\alpha.
\end{split}
} 
The difference between $E^S_0$ and $E^S_2$ is a $\mathcal{O}(1/\LL)$
correction, $E^S_0 - E^S_2 = 2\alpha/\LL$ and vanishes in the thermodynamic limit.

We now form normalized equal superpositions of $\psi'_{0,bb'}$ and
$\psi'_{2,bb'}$ states:
\beq{
\begin{split}
\Psi'_{0} &= \frac{1}{\sqrt{\LL(\LL-1)}} \sum_{b \neq b'} \psi'_{0,bb'},
\\
\Psi'_{2} &= \frac{1}{\sqrt{\LL(\LL-1)}} \sum_{b \neq b'} \psi'_{2,bb'},
\end{split}
}
we can be simplified in the $\LL\to\infty$ limit to
\beqz{
\begin{split}
\Psi'_{0} &\approx \frac{1}{\LL} \sum_{b \neq b'} c^\dag_{b\uparrow}
c^\dag_{b'\downarrow} \Psi^{\LL \backslash b,b'}_{M}, \\
\Psi'_{2} &\approx \frac{1}{\LL} \sum_{b \neq b'} c^\dag_{b\uparrow}
c^\dag_{b'\downarrow} \Psi^{\LL \backslash b,b'}_{M-1}, \\
\end{split}
}
These are spin-singlet states. 
Furthermore, we make a normalized spin-singlet combination of
$\psi_{1,b,\sigma}$:
\beq{
\begin{split}
\Psi_1 &= \frac{1}{\sqrt{\LL}} \sum_{b=1}^{\LL} \frac{1}{\sqrt{2}}  \left( \psi_{1,b,\uparrow} -
\psi_{1,b,\downarrow} \right) \\
&= \frac{1}{\sqrt{\LL}} \sum_{b=1}^{\LL} \frac{1}{\sqrt{2}}
\left( \ket{\uparrow} \otimes c^\dag_{b,\downarrow} - \ket{\downarrow}
\otimes c^\dag_{b,\uparrow} \right) \Psi^{\LL\backslash b}_{M}.
\end{split}
}
Making use of Eq.~\eqref{sqrtfactor} at half-filling ($M = \LL/2$), this can also be expressed as
\beq{
\begin{split}
\Psi_1 & \approx \left( \ket{\uparrow} \otimes f^\dag_\downarrow
- \ket{\downarrow} \otimes f^\dag_\uparrow \right) \Psi^{\LL}_{M}.
\end{split}
}

We note that no spin-triplet states are included here.

\begin{figure}
    \centering
    \includegraphics[width=8cm]{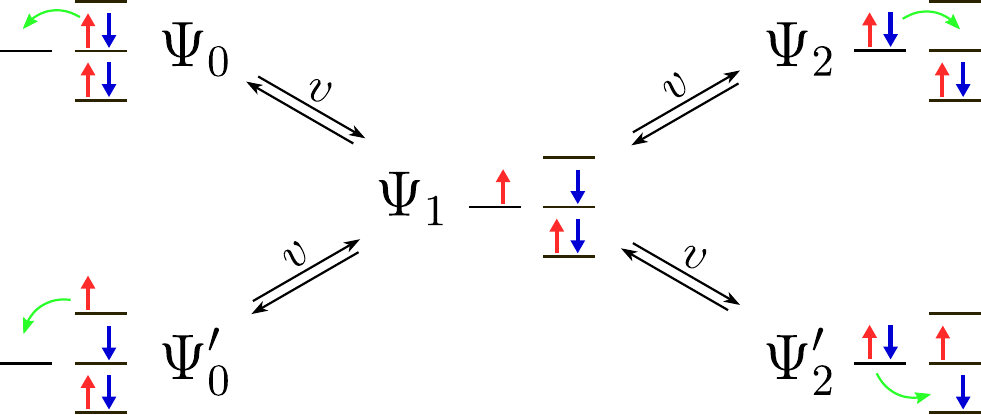}
    \caption{Schematic representation of the states coupled
    in the singlet subspace. In the flat-band limit, no other
    states couple with this subset.}
    \label{fig:S}
\end{figure}

To find the matrix elements, we make use of the result from
Appendix~\ref{app3} in the large $\LL$ limit:
\beq{
f^\dag_{\up} f^\dag_{\dn} \Psi^\LL_M \approx \frac{1}{2} 
\Psi^\LL_{M+1} + \frac{1}{2} \frac{1}{\LL}
\sum_{b\neq b'} c^\dag_{b\up} c^\dag_{b'\dn} \Psi^\LL_M.
}
We find, for example,
\beq{
\begin{split}
\langle \Psi_0 | f^\dag_\uparrow d_\uparrow | \Psi_1 \rangle
&= 
\bra{0} (\Psi^{\LL}_{M+1})^* (-d_\uparrow f^\dag_\uparrow) 
\ket{\uparrow} \otimes f_\downarrow^\dag \Psi^\LL_M  \\
&= \bra{0} d_\uparrow \ket{\uparrow}
(\Psi^{\LL}_{M+1})^* f^\dag_\uparrow
f_\downarrow^\dag \Psi^\LL_M \\
&\approx 1/2,
\end{split}
}
and similarly
\beq{
\begin{split}
\langle \Psi_0 | f^\dag_\downarrow d_\downarrow | \Psi_1 \rangle
\approx 1/2,
\end{split}
}
giving a total electron hopping matrix element of $1$. Other matrix
elements can be computed in an analogous way.
The effective Hamiltonian in the basis $\{ \Psi_1, \Psi_0,
\Psi'_0, \Psi_2, -\Psi'_2 \}$ in the large-$\LL$ limit is then
\beq{
\label{H2}
H^S_\mathrm{eff} = \begin{pmatrix}
E^S_1 & v      & v        & v      & v        \\
v     & E^S_0  & 0        & 0      & 0        \\
v     & 0      & {E^S_0}' & 0      & 0        \\ 
v     & 0      & 0        & E^S_2  & 0        \\
v     & 0      & 0        & 0      & {E^S_2}' \\
\end{pmatrix}.
}
The signs of basis states have been chosen so that the out-of-diagonal
matrix elements are all positive (in the sense of having the same sign as $v$).
The schematic representation of this Hamiltonian is shown in Fig.~\ref{fig:S}.
If the $1/\LL$ corrections in diagonal elements are neglected (as is
valid in the large $\LL$ limit), so that
\beq{
\begin{split}
E^S_{02}    &= -\alpha\frac{\LL}{2} + U/2,\\
{E^S_{02}}' &= -\alpha\frac{\LL}{2} + 2\alpha + U/2 = E^S_{02} + 2\alpha, \\
E^S_1       &= -\alpha\frac{\LL}{2} + \alpha, \\
\end{split}
}
the problem simplifies further to a $3\times3$ matrix problem
\beq{
\label{H2bis}
H^S_\mathrm{eff} = \begin{pmatrix}
E^S_1 & \sqrt{2} v         & \sqrt{2} v         \\
\sqrt{2} v     & E^S_{02}  & 0               \\
\sqrt{2} v     & 0         & {E^S_{02}}'        \\ 
\end{pmatrix}.
}

The numeric diagonalisation of this Hamiltonian produces a result which 
agrees very well with the DMRG results even for $\LL=50$, 
see Fig.~\ref{figU} (bottom panel). Exact matrix elements for any
values of $\LL$ and $M$ are provided in Appendix~\ref{appD}.

\subsection{Dependence on hybridisation strength}

We now consider the $\Gamma$-dependence of the singlet-doublet
excitation energy for different values of $U$. 

\subsubsection{Large-$U$ regime}

At the particle-hole symmetric point, the lowest eigenvalue of the
Hamiltonian $H^D_\mathrm{eff}$ in Eq.~\eqref{HD} is
\beq{
\label{EDX}
E_D = -\alpha\frac{\LL-1}{2} + \frac{U-\sqrt{16v^2+(U+2\alpha)^2}}{4}.
}
The expansion around the $U \to \infty$ limit is
\beq{
\begin{split}
E_D &\approx -\alpha\frac{\LL}{2} - 2 \frac{v^2}{U} + \frac{4 v^2 \alpha}{U^2} + \mathcal{O}(1/U^3) \\
&= -\alpha\frac{\LL}{2} - \frac{J}{4} + \frac{J \alpha}{2U} +
\mathcal{O}(1/U^3),
\end{split}
}
where $J=8v^2/U$ as defined in Eq.~\eqref{eq29}. The first term is the energy
of the half-filled decoupled bath, Eq.~\eqref{EE}. The second term is
the trivial linear shift of energies in the Schrieffer-Wolff
transformation \cite{schrieffer1966}, $-J/4$. The third term is the
leading order $J/U$ correction. As expected, it is proportional to
$\alpha$ and $J$, and inversely proportional to $U$ (the energy cost
of valence fluctuations), since it is due to the perturbation of the
SC state by the hybridisation with the impurity.

The lowest eigenvalue of the Hamiltonian $H^S_\mathrm{eff}$ in
Eq.~\eqref{H2bis} in the large-$U$ limit is
\beq{
\label{EU1}
\begin{split}
E_S &\approx -\alpha \frac{\LL-2}{2}-\frac{8v^2}{U} + \mathcal{O}(1/U^3) \\
&= -\alpha \frac{\LL-2}{2} - J + \mathcal{O}(1/U^3).
\end{split}
}
Now we consider the difference, i.e., the Yu-Shiba-Rusinov excitation
energy $E_\mathrm{YSR}=E_S-E_D$. At large $U$ we then recover the
Kondo-limit results from Sec.~\ref{kondo}, 
\beq{
E_\mathrm{YSR}=\alpha - \frac{3J}{4},
}
up to $1/\LL$ corrections.

For moderate $U$ one finds deviations from the simple
linear-in-$\Gamma$ behavior, see Fig.~\ref{figEYSR}. The YSR energy at
the same $\Gamma/U$ ratio is found to be {\it higher}, the lower $U$
is. The charge fluctuations thus reduce the effects of the exchange
coupling (at the same value of $J$). Note that at large $\Gamma$ the
eigenvalue reaches the limiting value $-(1-1/\LL)\alpha$ (``gap
edge'') with a finite slope, indicating a level crossing. This is the
main differenece compared with the problem with a finite bandwidth,
where the subgap state experiences a level repulsion and at large
$\Gamma$ asymptotically approaches the gap edge which is accompanied
by a continuous transfer of the spectral weight from the subgap states
to the continuum of Bogoliubov excitations \cite{hecht2008}. In the
flat-band case there is no continuum, only a set of degenerate
excitations which do not hybridize with the subgap state, thus the
hybridized state can cross this set unperturbed.
The nonlinearity in $\Gamma$ can be obtained by expanding the lowest 
eigenvalue of $H^S_\mathrm{eff}$ to the next order in $1/U$:
\beq{
\begin{split}
E_S &\approx -\alpha \frac{\LL-2}{2} - \frac{8v^2}{U}
+ \frac{128 v^4-32 v^2 \alpha^2}{U^3} \\
&= -\alpha \frac{\LL-2}{2} - \frac{8\Gamma}{\pi \rho U^2}
- \frac{32 \frac{\Gamma}{U} \left[ \left( \pi \rho \alpha \right)^2  -  4\pi \rho \frac{\Gamma}{U}
\right]}{(\pi \rho U)^3}.
\end{split}}

\subsubsection{Small-$U$ limit}

We now consider the expansion for small values of $U$. For
Eq.~\eqref{EDX} we find
\beq{
\label{mma1}
E_D \approx -\alpha \frac{\LL-1}{2} + \frac{U}{4}
\left( 1-\frac{\alpha}{\sqrt{\alpha^2+4v^2}} \right)
- \frac{\sqrt{\alpha^2+4v^2}}{2}.
}
Taking the small-$\Gamma$ limit gives $E_D \approx -\alpha \LL/2$,
thereby fully cancelling out the $U$ dependence.

\begin{figure}
\centering
\includegraphics[width=\columnwidth]{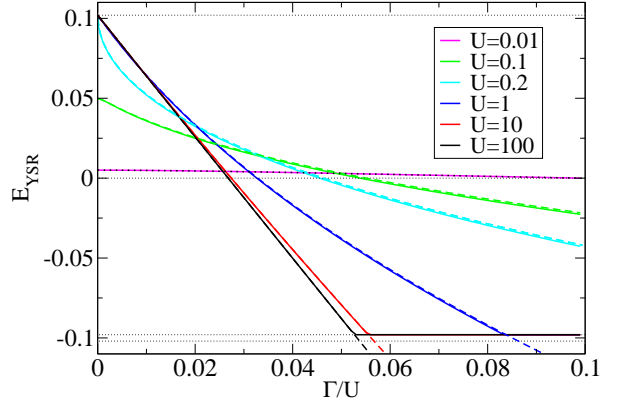}
\caption{Excitation energy $E_\mathrm{YSR}=E_S-E_D$ in the flat-band
limit for the $\Delta N=1$ transitions. Parameters are $\LL=50$,
$\alpha=0.1$. Full lines: DMRG results. Dashed lines: analytical
calculation. The small deviations are due to the $1/\LL$ correction that
for simplicity have not been included here.
}
\label{figEYSR}
\end{figure}

The lowest eigenvalue of the Hamiltonian $H^S_\mathrm{eff}$ in
Eq.~\eqref{H2bis} in the small-$U$ limit is
\beq{
\label{mma2}
E_S \approx -\alpha \frac{\LL-2}{2} + \frac{U}{2}
\frac{\alpha^2+2v^2}{\alpha^2+4v^2}
-\sqrt{\alpha^2+4v^2} .
}
Taking the $\Gamma\to0$ limit, we find $E_S \approx -\alpha
\LL/2 + U/2$. At zero $\Gamma$, the excitation energy is
\beq{
\Delta E(\Gamma=0)=\lim_{\Gamma\to0} \left( E_S-E_D \right) = U/2,
}
as expected and as seen in Fig.~\ref{figEYSR}. 
The lowest excitation is not a YSR singlet, but a ABS-like state with a QD in a proximitized state ($\ket{0} + \ket{2}$ superposition).
We note in passing that the order of taking the limits is important; here we took
the limit of $U\to0$ followed by $\Gamma\to0$.

The $\Gamma$ dependence to lowest order is
\beq{
\Delta E = E_S-E_D \approx \frac{U}{2}
- \frac{\Gamma}{\pi \rho \alpha} \left(
1+ \frac{3U}{2\alpha} \right).
}
The energy reduction in the Andreev-bound-state regime is controlled
by the $\Gamma/\alpha$ ratio, a non-zero repulsion $U$ is merely an
order $U/\alpha$ correction to the prefactor.

\subsubsection{Special case of $U=2\alpha$}

In the limit of $U=2\alpha$, some states in the singlet subspace
become degenerate: $E^S_{02} = E^S_1$. In this regime,
we find the following expansions in the weak hybridisation limit:
\beq{
\begin{split}
E_D &= -\alpha \frac{\LL}{2} - \frac{v^2}{2\alpha}, \\
E_S &= -\alpha \frac{\LL}{2} - \frac{v^2}{2\alpha} + \alpha - \sqrt{2} v.
\end{split}
}
This leads to the interesting result that
\beq{
E_S-E_D = \alpha - \sqrt{ \frac{2\Gamma}{\pi \rho} }.
}
This square root singularity is indeed observed in the results shown
in Fig.~\ref{figEYSR}.

The singular behaviour observed at $U=2\alpha$ marks
the boundary between the regimes of proximitized (Andreev)
subgap states for $U<2\alpha$ and the Yu-Shiba-Rusinov states
for $U>2\alpha$. This is a nontrivial observation because
it is generally believed that the cross-over between
these two limits is perfectly smooth. In fact, there exists a clear anomaly at $U=2\alpha$ 
in the low-$\Gamma$ limit.

\subsection{Dependence on gate voltage}

\begin{figure}
\centering
\includegraphics[width=\columnwidth]{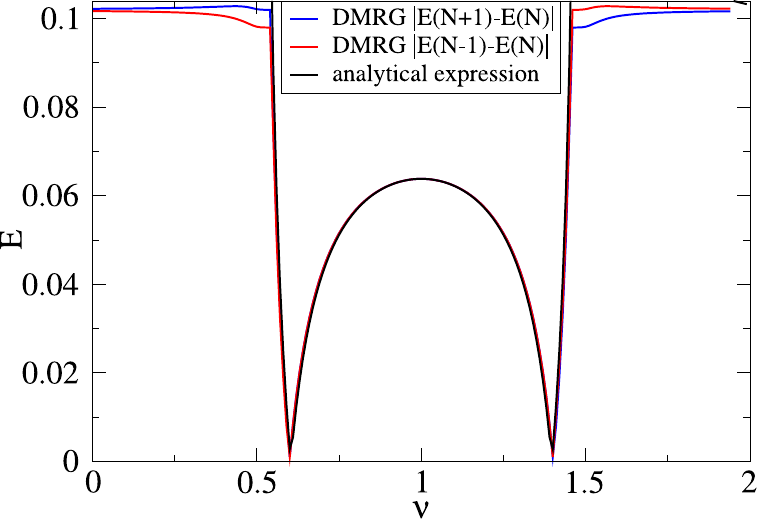}
\caption{YSR subgap state dispersion (positive frequency branch) in the flat-band limit. 
The lineshape corresponds to the conventional YSR loops.
Parameters are $U=100$, $\LL=50$, $\Gamma/U=0.01$, $\epsQD=U(1/2-\nu)$, where
$\nu$ controls the impurity filling.}
\label{figph}
\end{figure}

The formalism also correctly describes the transitions to different
impurity occupancies when $\epsQD=U(1/2-\nu)$ with $\nu\neq1$. We find good agreement with the reference 
DMRG results, see Fig.~\ref{figph}.
This is the case even taking the truncated 3-dimensional basis from Eq.~\eqref{HD}.
For
$\nu$ sufficiently far away from $1$, the doublet solution enters the
quasicontinuum: the black line in Fig.~\ref{figph} shoots past
the gap edge $\alpha(1+1/\LL)$. The true lowest excited state is
then a decoupled Bogoliubov quasiparticle sitting at the bottom of the
quasicontinuum.

In the large-$U$ limit, the lowest lying eigenstates are
\beq{
\begin{split}
E_S &\approx -\alpha \frac{\LL-2}{2} + \frac{8v^2}{u(3+4 \nu
(\nu-2)}, \\
E_D &\approx -\alpha \frac{\LL}{2} + \frac{2v^2}{u(3+4\nu
(\nu-2)},
\end{split}
}
so that 
\beq{
\begin{split}
\Delta E = E_S-E_D &= \frac{(\LL+1)}{\LL} \alpha +
\frac{6v^2}{3+4\nu(\nu-2)} \\
&\approx \alpha - \frac{3J}{4},
\end{split}
}
where we noted that away from half filling, the
Kondo exchange coupling is
\beq{
\begin{split}
J &= \frac{2v^2}{-\epsilon} + \frac{2v^2}{U+\epsilon}\\
&= \frac{1}{4\nu(2-\nu)-3}\frac{8v^2}{U}.
\end{split}
}
As expected, to first order in $1/U$, the only effect of departure
from the particle-hole asymmetry is the modified value of the Kondo
exchange. The leading correction is of order $\alpha J/U$, as found
in the previous section.

\section{Comparison to the BCS zero-bandwidth approximation}
\label{ZBW}

A commonly used approximation in QD-SC models is the ZBW limit of BCS, 
where the SC is modelled by a single level with mean-field pairing interaction \cite{affleck2000,vecino2003,bergeret2007,ysrdqds}. 
Here we show that this description is equivalent to the RM in the $\gamma \to 0$ limit.  
Consider the ZBW Hamiltonian
\beq{
H = 
 \epsQD (\hat{n}-1) + \UQD\hat{n}_\uparrow \hat{n}_\downarrow 
+ v \sum_{i,\sigma} \left( f^\dag_\sigma
d_\sigma + d^\dag_\sigma f_{\sigma} \right)
- \Delta (f^\dag_\up f^\dag_\dn + \text{H.c.} ).
}
In the doublet subspace with the basis states $\{ d^\dag_\up,
d^\dag_\up f^\dag_\up f^\dag_\dn, f^\dag_\up, d^\dag_\up d^\dag_\dn
f^\dag_\up \}$ (applied to the zero occupancy vacuum state) one finds
the following matrix representation:
\newcommand{\HDZ}{H^D_\mathrm{ZBW}}
\newcommand{\HDe}{H^D_\mathrm{eff}}
\beq{
\HDZ = \begin{pmatrix}
0 & -\Delta & v & 0 \\
-\Delta & 0 & 0 & -v \\
v & 0 & -\epsilon & 0 \\
0 & -v & 0 & U+\epsilon 
\end{pmatrix}.
}
Alternatively, one may work in the basis 
\beqz{
\begin{split}
& \left(d^\dag_\up + d^\dag_\up f^\dag_\up f^\dag_\dn\right)/\sqrt{2} = d^\dag_\up  \left(1 + f^\dag_\up f^\dag_\dn\right)/\sqrt{2}, \\
& \left(d^\dag_\up - d^\dag_\up f^\dag_\up f^\dag_\dn\right)/\sqrt{2} = d^\dag_\up  \left(1 - f^\dag_\up f^\dag_\dn\right)/\sqrt{2},\\
& f^\dag_\up, \\
& d^\dag_\up d^\dag_\dn f^\dag_\up,
\end{split}
}
where the matrix representation is:
\beq{
\HDZ = \begin{pmatrix}
-\Delta & 0  & v/\sqrt{2} & -v/\sqrt{2} \\
0 & +\Delta & v/\sqrt{2} & v/\sqrt{2} \\
v/\sqrt{2} & v/\sqrt{2} & -\epsilon & 0 \\
-v/\sqrt{2} & v/\sqrt{2} & 0 & U+\epsilon 
\end{pmatrix},
}
which is equivalent to $\HDe$ in Eq.~\eqref{HDeff} up to a trivial shift of all diagonal elements.
The combination $(1 + f^\dag_\up f^\dag_\dn)/\sqrt{2}$ represents a Cooper-pair state,
while the combination $(1 - f^\dag_\up f^\dag_\dn)/\sqrt{2}$ represents a two-quasiparticles state with the energy cost of $2\Delta$.

In the singlet subspace, we work in the basis (the signs are chosen so as to produce positive
signs for the out-of-diagonal matrix elements)
\beqz{
\begin{split}
&(d^\dag_\up f^\dag_\dn - d^\dag_\dn f^\dag_\up)/\sqrt{2},\\
&\left(1 + f^\dag_\up f^\dag_\dn\right)/\sqrt{2},\\
&-\left(1 - f^\dag_\up f^\dag_\dn\right)/\sqrt{2},\\
&d^\dag_\up d^\dag_\dn \left(1 + f^\dag_\up f^\dag_\dn\right)/\sqrt{2},\\
&d^\dag_\up d^\dag_\dn \left(1 - f^\dag_\up f^\dag_\dn \right)/\sqrt{2},
\end{split}
}
where the matrix takes the following form
\newcommand{\HSZ}{H^S_\mathrm{ZBW}}
\newcommand{\HSe}{H^S_\mathrm{eff}}
\beq{
\HSZ = \begin{pmatrix}
0 & v & v & v & v \\
v & -\Delta-\epsilon & 0 & 0 & 0 \\
v & 0 & \Delta-\epsilon & 0 & 0 \\
v & 0 & 0 & -\Delta+U+\epsilon & 0 \\
v & 0 & 0 &  & \Delta+U+\epsilon
\end{pmatrix}.
}
Up to a trivial energy shift, this is exactly the same as $\HSe$ from Eq.~\eqref{H2}.

The equivalence between the models is exact in the ZBW limit $\gamma \to 0$, when $\alpha$ plays the role of excitation energy gap, akin to $\Delta$ in BCS.
The demonstration of this equivalence is one of the key results of this work.

\section{Conclusion}
\label{discuss}

Even though superconductivity and Bose-Einstein condensation are apparently quite different states of matter, they emerge as a result of the same attractive pairing interaction in the opposite limits of pairing strength. When a SC is coupled to a magnetic impurity, it induces discrete singlet states inside the SC gap. 
We find that the subgap states persist throughout the weak to strong pairing crossover, and we furthermore show that their nature remains qualitatively the same. 
The differences are limited to the scaling of hybridization strength and excitation gap, both due to the decreasing importance of bath kinetic energy with increased pairing.

In the limit of strong pairing interaction the kinetic energy term can be omitted. The single-particle levels in the bath are then degenerate, which lets us obtain a flat-band Hamiltonian in the reduced Hilbert space spanned by just a few many-particle states connected by the Hamiltonian. The BEC limit of the problem is in this sense analytically solvable, and we show that the results agree very well with exact calculations. We also show that the flat-band Hamiltonian in the limit of large system is formally exactly equivalent to the BCS zero bandwidth approximation. 

ZBW approximations and similar toy models are very commonly applied when modeling QDs coupled to SCs \cite{yosida1966, ysrdqds, Zitko2017, vonOppen2021, SchmidYSRdimer2022}. Our result explains their surprisingly strong predictive power and thus provides a solid foundation for their application in this context.
Typically the argument for using the ZBW approximation is that a single level at the gap edge is sufficient to encompass the important phenomena, because the SC has a diverging density of states at the gap edge and because these near-edge are the most important for the subgap state formation: in fact, the subgap state weight originates precisely from the depletion of the spectral weight from the superconducting coherence peaks.
The higher excited states largely remain unperturbed by the impurity, and their small contributions can be discarded. We
find that this is a very reasonable approximation in a variety of situations.

Due to its simplicity, the flat-band model can be straightforwardly extended in various ways. 
To accurately describe mesoscopic SC islands, it is important to take into account the Coulomb repulsion between the island electrons \cite{coulomb2}. This can be done by augmenting the model with a charging energy term $E_C \hat{n}_\mathrm{SC}^2$. 
Because the RM conserves particle number we are operating in the canonical ensemble, thus implementing such operator is very simple and it is straightforward to extend the effective flat-band model by adding $E_C$ to appropriate diagonal matrix elements.
For a problem with a single SC bath, this merely reduces the energy scale for charge fluctuations from $U/2$ to $U/2+E_c$ \cite{coulomb1,coulomb2,coulomb3}.
The Zeeman splitting can also be trivially included in the flat-band model. Additional triplet (and quadruplet) states that become relevant at large magnetic fields can also be included in consideration. 

The analytical approach can also be extended to multichannel impurity problems. These have additional degrees of freedom, namely the occupation number differences between channels (i.e. the distribution of Cooper pairs between the two superconductors). 
An appropriate linear combination of such states corresponds to well defined phase differences which would appear in the equivalent BCS formalism, while a charging term would induce phase fluctuations. We will
pursue this direction in our future work.

\begin{acknowledgments}
We acknowledge discussions with Gorm Steffensson, Jens Paaske, and
Martin \v{Z}onda. We acknowledge the support of the Slovenian Research
Agency (ARRS) under P1-0044, P1-0416, and J1-3008.
\end{acknowledgments}


\appendix

\section{Eigenenergy calculation}
\label{app1}

Summing Eq.~\eqref{com1} over $i,j$ in the set $\UU$ we obtain
\beq{
[B,B^\dag] = 1-\frac{1}{\UU} \hat{N},
\label{com2}
}
where $\hat{N}=\sum_i^\UU \hat{n}_i$ is the total number of electrons in
the unblocked levels.

In this appendix we evaluate $B^\dag B$ on state $\ket{M}=(B^\dag)^M \ket{0}$.
By repeatedly commuting $B^\dag$ and $B$ we find
\beq{
\label{eq14}
\begin{split}
&B^\dag B \ket{M} =\left( B^\dag B \right) (B^\dag)^M  \ket{0} \\
&= \left[ B^\dag \left(B^\dag B + 1 - \frac{1}{\UU} \hat{N} \right)
(B^\dag)^{M-1} \right] \ket{0} \\
&= \left[
(B^\dag)^2 B (B^\dag)^{M-1}
+ \left( 1- \frac{2(M-1)}{\UU} \right) (B^\dag)^M 
\right] \ket{0} \\
&= \Big[
(B^\dag)^3 B (B^\dag)^{M-2}
+ \left( 1- \frac{2(M-2)}{\UU} \right) (B^\dag)^M \\ 
& \quad + \left( 1- \frac{2(M-1)}{\UU} \right) (B^\dag)^M 
\Bigr] \ket{0} \\
&= \left[ \sum_{m=1}^M \left( 1 - \frac{2(m-1)}{\UU} \right) \right]
\ket{M} = c_M \ket{M},
\end{split}
}
with
\beq{
c_M = \frac{(1+\UU-M)M}{\UU}.
}

\section{Action of creation operators on $\Psi^\LL_M$}
\label{app3}

In this Appendix we discuss the action of fermionic operators
on the eigenstates $\Psi^{\LL,\UU}_M$ of the Richardson's model.
This defines the relations between the states from subspaces
distinguished by the different number $M$ of Cooper pairs. 
The subsectors differing by $\Delta M = \pm 1$
are coupled due to the Cooper-pair-breaking property of the
exchange interaction induced by the presence of the magnetic impurity.
The relations derived in the following provide the full set
of algebraic relations required to determine the coupling
between the accessible states and all matrix elements
in the flat band limit.

\subsection{Single-quasiparticle state}

In this section the number of levels is fixed to $\LL$. When a single
superscript is indicated for brevity, the number indicates the number of unblocked levels,
i.e., $\Psi^{\UU}_M \equiv \Psi^{\LL,\UU}_M$. When no superscript is indicated,
all levels are unblocked, i.e., $\Psi_M \equiv \Psi^{\LL,\LL}_M$.

Let us consider the spin-doublet state
\beq{
\psi_{\sigma} = f^\dag_\sigma \Psi^{\LL,\LL}_M = \frac{1}{\sqrt{\LL}}
\sum_{b=1}^\LL c^\dag_{b,\sigma} \Psi^{\LL,\LL}_M
}
with $\Psi^{\LL,\LL}_M$ being the normalized eigenstate for $M$ Cooper
pairs in a system with $\LL$ levels, all of them unblocked. The
application of the creation operator creates a superposition of states
with one blocked level indexed by $b$. This state is thus spanned by
the following {\it orthonormal} set of $\LL$ eigenstates of the SC
bath:
\beq{
\phi_{b\sigma} = c^\dag_{b,\sigma} \Psi^{\LL,\LL\backslash b}_M,
}
where $\LL\backslash b$ in the superscript of $\Psi$ indicates that the level $b$ is
blocked. If all levels are equal, so are the coefficients, thus
\beq{
\psi_{\sigma} = C \frac{1}{\sqrt{\LL}} \sum_{b=1}^\LL \phi_{b\sigma}.
}
Let us now fix the proportionality constant $C$:
\beq{
\braket{\psi_\sigma}{\psi_\sigma} 
= \bra{\Psi^\LL_M} f_\sigma f^\dag_\sigma \ket{\Psi^\LL_M}
= 1 - \bra{\Psi^\LL_M} f^\dag_\sigma f_\sigma \ket{\Psi^\LL_M} =
C^2.
}
We expand in the site basis:
\beq{
\begin{split}
f^\dag_\sigma f_\sigma &= \frac{1}{\LL} \sum_{i,j} c^\dag_{i,\sigma}
c_{j,\sigma} \\
&= \frac{1}{\LL} \sum_i \hat{n}_{i,\sigma} + \frac{1}{\LL} \sum_{i \neq j} c^\dag_{i,\sigma}
c_{j,\sigma} \\
&= \frac{\hat{N}_\sigma}{\LL} + \frac{1}{\LL} \sum_{i \neq j} c^\dag_{i,\sigma}
c_{j,\sigma}.
\end{split}
}
The first term gives $M/\LL$. The second gives zero, because when
applied to the ket it produces a linear combination of terms which all
contain singly occupied levels, and such a state is orthogonal to
$\Psi^\LL_M$ in the bra. Finally 
\beq{
C^2 = 1 - \frac{M}{\LL} = \frac{\LL-M}{\LL}.
}
We conclude that
\beq{
\psi_\sigma =
f^\dag_\sigma \Psi^{\LL,\LL}_{M} = \sqrt{ \frac{\LL-M}{\LL} }
\frac{1}{\sqrt{\LL}} \sum_b
c^\dag_{b,\sigma} \Psi^{\LL,\LL\backslash b}_{M}.
}
This expression is exact.
For $M\to0$, in the absence of any Cooper pairs, the
prefactor becomes equal to 1, since the electron can be added to
any level. For $M\to \LL$, the prefactor tends to 0, since there
are only a few levels available to accommodate the additional electron.
Finally, for half filling the prefactor becomes $1/\sqrt{2}$, thus
\beq{
\psi_\sigma =
f^\dag_\sigma \Psi^{\LL,\LL}_{\LL/2} = \frac{1}{\sqrt{2}}
\frac{1}{\sqrt{\LL}} \sum_b
c^\dag_{b,\sigma} \Psi^{\LL\backslash b}_{\LL/2}.
}
Note that this state $\psi_\sigma$ is {\it not} normalized to 1.

In the same way one derives
\beq{
\psi'_\sigma =
f_{\bar \sigma} \Psi^{\LL,\LL}_{M} = (-1)^\sigma \sqrt{ \frac{M}{\LL} }
\frac{1}{\sqrt{\LL}} \sum_b
c^\dag_{b,\sigma} \Psi^{\LL,\LL\backslash b}_{M-1},
}
where $(-1)^\uparrow=-1$ and $(-1)^\downarrow=1$. This expression
is exact.
It simplifies at half filling to 
\beq{
\psi'_\sigma = (-1)^\sigma
f_{\bar \sigma} \Psi^{\LL,\LL}_{\LL/2} = \frac{1}{\sqrt{2}}
\frac{1}{\sqrt{\LL}} \sum_b
c^\dag_{b,\sigma} \Psi^{\LL,\LL\backslash b}_{\LL/2-1},
}
The sign factor is due to the order convention of spin-up and
spin-down operators in the hard-core boson operators $A_i$.

\subsection{Two-quasiparticle state}

The other state of interest is the spin-singlet state
\beq{
\begin{split}
\psi_{2} &= f^\dag_{\up} f^\dag_{\dn} \Psi^\LL_M \\
&= \frac{1}{\LL} \sum_{i,j} c^\dag_{i\up} c^\dag_{j\dn} \Psi^\LL_M \\
&= \frac{1}{\LL} \sum_i A^\dag_i \Psi^\LL_M
+ \frac{1}{\LL} \sum_{b \neq b'} c^\dag_{b\up} c^\dag_{b'\dn} \Psi^\LL_M \\
&= \frac{1}{\sqrt{\LL}} B^\dag \Psi^\LL_M
+ \frac{1}{\LL} \sum_{b \neq b'} c^\dag_{b\up} c^\dag_{b'\dn} \Psi^\LL_M.
\end{split}
}
We have
\beq{
B^\dag \Psi^\LL_M = B^\dag \NN_M \ket{M} =
\NN_M \ket{M+1} = \frac{\NN_M}{\NN_{M+1}} \Psi^\LL_{M+1}.
}
The normalization factors $\NN_M$ are evaluated in
Appendix~\ref{norm}. The states with two blocked levels are spanned by
the {\it orthonormal} set of $\LL^2-\LL=\LL(\LL-1)$ states (for $b\neq
b'$)
\beq{
\phi_{bb'} = c^\dag_{b,\up} c^\dag_{b',\dn} \Psi^{\LL,\LL\backslash b,b'}_M.
}
We thus have
\beq{
\frac{1}{\LL}  \sum_{b \neq b'} c^\dag_{b\up} c^\dag_{b'\dn}
\Psi^\LL_M = C \frac{1}{\sqrt{\LL(\LL-1)}} \sum_{b \neq b'} \phi_{bb'},
}
with a proportionality constant $C$ that needs to be determined.
For that purpose, we compute the norm:
\beq{
C^2 = \frac{1}{\LL^2}
\sum_{b\neq b', c \neq c'}
\bra{ \Psi^\LL_M} c_{c'\dn} c_{c\up} c_{b\up}^\dag c_{b'\dn}^\dag
\ket{\Psi^\LL_M}.
}
Only the terms with $c=b$ and $c'=b'$ contribute, otherwise the states are
orthogonal. 
Now
\beq{
c_{b'\dn} c_{b\up} c_{b\up}^\dag c_{b'\dn}^\dag
= 1-\hat{n}_{b\up} - \hat{n}_{b'\dn}+\hat{n}_{b\up} \hat{n}_{b'\dn}.
}
We need to handle the restriction to $b\neq b'$ carefully,
noting that $\sum_{b \neq b'} 1 = \LL(\LL-1)$,
$\sum_{b \neq b'} \hat{n}_{b\sigma} = N_\sigma (\LL-1)$,
and
\beq{
\sum_{b \neq b'} \hat{n}_{b\up} \hat{n}_{b'\dn} =
\sum_{b,b'} \hat{n}_{b\up} \hat{n}_{b'\dn} - \sum_b \hat{n}_{b\up}
\hat{n}_{b\dn}
= \hat{N}_\up \hat{N}_\dn - \sum_b \hat{P}_{2,b}.
}
\begin{widetext}
The operator $\hat{P}_2 = \sum_b \hat{P}_{2,b}$ counts the number of
pairs in the system, therefore $\bra{\Psi_M} \hat{P}_2 \ket{\Psi_M} =
M$. Collecting all terms we then find
\beq{
C^2 = \frac{1}{\LL^2}
\bra{\Psi^\LL_M}
\left( \LL(\LL-1) - (\LL-1) \hat{N}_\up - (\LL-1) \hat{N}_\dn + \hat{N}_\up
\hat{N}_\dn - \hat{P}_2 \right) \ket{\Psi^\LL_M} = \frac{(\LL-M)(\LL-M-1)}{\LL^2}.
}
For $M=0$, the factor $C^2 = 1-\frac{1}{\LL}$, which goes to 1 in the
large-$\LL$ limit. In the opposite limit of $M=\LL$, it is zero.
Finally, for half filling, it tends to $C^2=1/2$ in the large-$\LL$
limit.

We write $\psi_2$ in the form
\beq{
\psi_2 =  f^\dag_{\up} f^\dag_{\dn} \Psi^\LL_M = \frac{\sqrt{(1+M)(\LL-M)}}{\LL} \Psi^\LL_{M+1} +
\frac{\sqrt{(\LL-M)(\LL-M-1)}}{\LL} \frac{1}{\sqrt{\LL(\LL-1)}} \sum_{b\neq b'} \phi_{bb'}.
}
This expression is exact.
There is an overall $\sqrt{\LL-M}$ factor. The relative importance of
the first and the second term scale as $\sqrt{M}$ and $\sqrt{\LL-M}$,
respectively, up to $1/\LL$ corrections.

For $M=\LL/2$, we finally find
\beq{
\psi_2 = f^\dag_{\up} f^\dag_{\dn} \Psi^\LL_M = \frac{1}{2}
\sqrt{\frac{2+\LL}{\LL}} \Psi^\LL_{M+1} 
+ \frac{1}{2} \sqrt{\frac{\LL-2}{\LL-1}} \frac{1}{\LL} \sum_{b\neq b'} c^\dag_{b\up} c^\dag_{b'\dn} \Psi^{\LL\backslash b,b'}_M.
}
Note that for large $\LL$ both square roots tend towards 1.
The states with two blocked levels are orthogonal to $\Psi^\LL_{M+1}$ states. Hence
\beq{
\braket{\psi_2}{\psi_2} 
= \frac{(1+M)(\LL-M)}{\LL^2}  + \frac{(\LL-M)(\LL-M-1)}{\LL^2} 
= \frac{\LL-M}{\LL}.
}
For $M=\LL/2$, this gives
\beq{
\braket{\psi_2}{\psi_2} = \frac{1}{2}.
}
The state $\psi_2=f^\dag_\uparrow f^\dag_\downarrow \Psi^\LL_M$ is thus not normalized to 1.

In an analogous way one calculates
\beq{
\psi'_2 =  f_{\dn} f_{\up} \Psi^\LL_M =
\frac{\sqrt{M(1+\LL-M)}}{\LL} \Psi^\LL_{M-1} -
\frac{\sqrt{M(M-1)}}{\LL} \frac{1}{\sqrt{\LL(\LL-1)}}
\sum_{b\neq b'} \phi'_{bb'},
}
with $\phi'_{bb'} = c^\dag_{b,\up} c^\dag_{b',\dn}
\Psi^{\LL\backslash b,b'}_{M-2}$. This expression is exact.
At half filling, it simplifies to
\beq{
\psi'_2 =  f_{\dn} f_{\up} \Psi^\LL_{\LL/2} =
\frac{1}{2} \sqrt{ \frac{2+\LL}{\LL} } \Psi^\LL_{\LL/2-1} -
\frac{1}{2} \sqrt{ \frac{\LL-2}{\LL-1} } \frac{1}{\LL}
\sum_{b\neq b'} \phi'_{bb'},
}
with the same prefactors as for $\psi_2$.

\end{widetext}

\section{Evaluation of normalization $\NN_M$}
\label{norm}

We fix $\NN_M$ from the condition
\beq{
\langle \Psi_M | \Psi_M \rangle = 
\NN_M^2 \langle M | M \rangle = 1.
}
We write 
\beq{
d_M = \langle M | M \rangle = \langle 0 | B^M (B^\dag)^M | 0 \rangle,
}
so that $\NN_M = (d_M)^{-1/2}$. Clearly $d_1=1$ and $\NN_1=1$.

Recall Eq.~\eqref{eq14}, $B^\dag B \ket{M} = c_M \ket{M}$. This
can also be expressed as $B \ket{M} = c_M \ket{M-1}$.
Thus we find
\beq{
\begin{split}
d_M &= \langle M | M \rangle \\
&= \langle M-1 | B | M \rangle \\
&= c_M \langle M-1 | M-1 \rangle \\
&= c_M d_{M-1}.
\end{split}
}
The solution of this recursion equation is
\beq{
d_M = (-\UU)^{-M} (1)_M (-\UU)_M,
}
where $(a)_n = \Gamma(a+n)/\Gamma(a)$ is the Pochhammer symbol.
At half filling $\UU=\LL$ and $M=\LL/2$, we find
\beq{
d_{\LL/2} = \frac{\LL!}{\sqrt{\LL^\LL}}.
}
Furthermore, for $\UU=\LL$ one has
\beq{
\frac{\NN^2_{M}}{\NN^2_{M+1}} = \frac{d_{M+1}}{d_M} 
= \frac{(1+M)(\LL-M)}{\LL}.
}

\section{Reference calculation}
\label{appD}

The method presented in Sec.~\ref{exact} is exact, but for clarity we have presented in that section only simplified results obtained
in the $\LL\to\infty$ limit at half-filling. Below are the exact expressions for $H^D_\mathrm{eff}$
and $H^S_\mathrm{eff}$.
\begin{widetext}
\beq{
\label{HDeffnew}
H^D_\mathrm{eff} = 
\begin{pmatrix}
    E^D_1   & 0         & v \sqrt{\frac{\LL - M}{\LL}}  &   -v \sqrt{\frac{M}{\LL}} \\
    0       &  E^{D'}_1 & v \sqrt{\frac{M}{\LL}}         & v \sqrt{\frac{\LL - M}{\LL}} \\
    v \sqrt{\frac{\LL - M}{\LL}} & v \sqrt{\frac{M}{\LL}} & E^D_0 & 0 \\
    -v \sqrt{\frac{M}{\LL}} & v \sqrt{\frac{\LL - M}{\LL}} & 0   & E^D_2 
\end{pmatrix}.
}

\beq{
\label{H2new}
H^S_\mathrm{eff} = \begin{pmatrix}
E^S_1 & v \sqrt{\frac{2(M + 1)}{\LL}} & v \sqrt{ \frac{2(\LL - M - 1)}{\LL}} & v \sqrt{ \frac{2(\LL - M)}{\LL}} & v \sqrt{ \frac{2M}{\LL}}        \\
v \sqrt{\frac{2(M + 1)}{\LL}}        & E^S_0  & 0        & 0      & 0        \\
v \sqrt{ \frac{2(\LL - M - 1)}{\LL}}  & 0      & {E^S_0}' & 0      & 0        \\ 
v \sqrt{ \frac{2(\LL - M)}{\LL}}      & 0      & 0        & E^S_2  & 0        \\
v \sqrt{ \frac{2M}{\LL}}            & 0      & 0        & 0      & {E^S_2}' \\
\end{pmatrix}.
}
\end{widetext}

We provide a Mathematica notebook with these definitions, as well as a corresponding input file for the DMRG solver, as Supplemental material \cite{SM}.

\bibliography{main}

\end{document}